\documentclass[twocolumn,superscriptaddress,preprintnumbers,amsmath,amssymb]{revtex4}
\usepackage{graphicx}
\usepackage{dcolumn}
\usepackage{bm}
\usepackage{hyperref}
\usepackage{amssymb,amsmath,amsthm,amsfonts}
\usepackage{bbm}
\usepackage{mathtools}
\usepackage{enumitem}
\usepackage{color}
\usepackage{textcomp}
\usepackage{gensymb} 
\usepackage{float}
\usepackage{soul}
\usepackage{booktabs}
\usepackage{graphicx}
\usepackage{siunitx}
\usepackage{multirow}
\usepackage{makecell}
\usepackage{draftwatermark}

\newcommand{\eq}[1]{(\ref{eq:#1})}
\newcommand{\fig}[1]{\hyperref[fig:#1]{Fig.~\ref*{fig:#1}}}
\newcommand{\figs}[1]{\hyperref[fig:#1]{Figs.~\ref*{fig:#1}}}
\newcommand{\Fig}[1]{\hyperref[fig:#1]{Figure~\ref*{fig:#1}}}
\newcommand{\Figs}[1]{\hyperref[fig:#1]{Figures~\ref*{fig:#1}}}
\newcommand{\tab}[1]{\hyperref[tab:#1]{Table~\ref*{tab:#1}}}
\newcommand{\routine}[1]{\hyperref[#1]{Routine~\ref*{#1}}}
\global\long\def\l({\left(}
\global\long\def\r){\right)}

\newcommand{\Gate}[1]{\textsc{#1}}

\newcommand{\tgate}{\Gate{t}}

\newcommand{\cnotgate}{\Gate{cnot}}
\newcommand{\rzgate}{{\Gate{r}}_{z}}

\newcommand{\xxgate}{{\Gate{xx}}}
\newcommand{\xphigate}{{\Gate{x$\phi$}}}
\newcommand{\xmphigate}{{\Gate{x$\bar{\phi}$}}}

\makeatletter
\newcommand{\vast}{\bBigg@{4}}
\newcommand{\Vast}{\bBigg@{5}}
\newcommand{\VVast}{\bBigg@{11.8}}
\makeatother
\begin{document}

\title{Power-optimal, stabilized entangling gate between trapped-ion qubits}

\author{Reinhold Bl\"umel}
\email{blumel@ionq.co}
\affiliation{IonQ, College Park, MD 20740, USA}
\affiliation{Wesleyan University, Middletown, CT 06459, USA}
\author{Nikodem Grzesiak}
\email{grzesiak@ionq.co}
\affiliation{IonQ, College Park, MD 20740, USA}
\author{Neal Pisenti}
\affiliation{IonQ, College Park, MD 20740, USA}
\author{Kenneth Wright}
\affiliation{IonQ, College Park, MD 20740, USA}
\author{Yunseong Nam}
\email{nam@ionq.co}
\affiliation{IonQ, College Park, MD 20740, USA}
\affiliation{Department of Physics, University of Maryland, College Park, MD 20742, USA}

\begin{abstract} 

To achieve scalable quantum computing, improving entangling-gate fidelity and its implementation-efficiency are of utmost importance. We present here a linear method to construct provably power-optimal entangling gates on an arbitrary pair of qubits on a trapped-ion quantum computer. This method leverages simultaneous modulation of amplitude, frequency, and phase of the beams that illuminate the ions and, unlike the state of the art, does not require any search in the parameter space. The linear method is extensible, enabling stabilization against external parameter fluctuations to an arbitrary order at a cost linear in the order. We implement and demonstrate the power-optimal, stabilized gate on a trapped-ion quantum computer.

\end{abstract}

\maketitle
Representing and processing information according to the laws 
of quantum physics, a quantum computer 
may surpass the computational power of a classical computer by many orders of magnitude, and
is expected to transform areas such as machine learning \cite{ar:HHL,ar:GenMod}, cryptosystems \cite{ar:Shor}, materials science \cite{ar:NF,ar:HINT}, and finance \cite{finance1,finance2}, to name only a few. Improving the reliability of quantum computation beyond the level of today's machines \cite{UMD,IonQ,Solidstate} is therefore critical to promote the quantum computer from a subject of academic interest to a powerful tool for solving problems of practical importance and utility.

The trapped-ion quantum information processor (TIQIP) 
is one of the most promising architectures for achieving a universal, programmable quantum computer, operating according to the gate model of quantum computing. Apart from a set of single-qubit gates, only a single entangling, two-qubit gate is necessary for achieving this goal. Today's TIQIPs \cite{UMD,IonQ} typically use an Ising $\xxgate$ gate, following the M{\o}lmer-S{\o}rensen protocol \cite{MS-1,AM,AM2}, as the two-qubit native gate. Its best reported fidelity is $99.9\%$ \cite{Gaebler2016,Ballance2016}, which may be compared with the best reported fidelity $99.9999\%$ of single-qubit gates \cite{Harty2014}. A host of  pulse-shaping techniques have been devised \cite{UMD,AM,AM2,FM,PM} to better control the underlying trapped-ion quantum systems for more efficient $\xxgate$ gate implementation, while reducing errors.

\begin{figure}
\centering
\includegraphics[scale=0.65,angle=0]{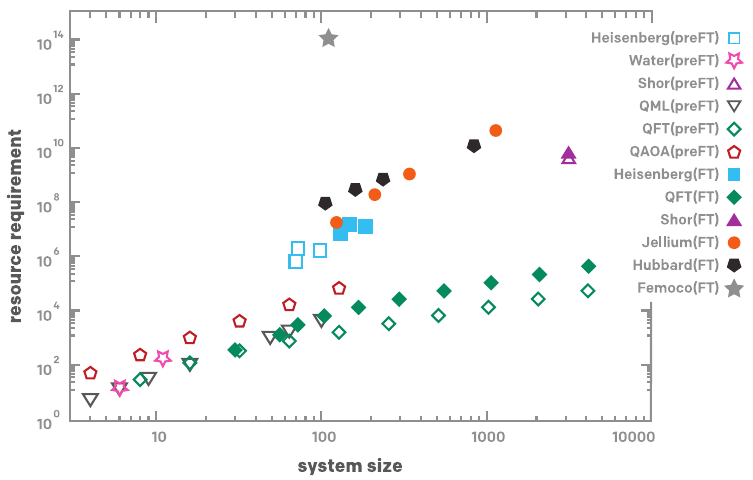}
\caption{Resource requirement of various quantum circuits as a function of system size. For the pre-FT regime, the resource cost is measured in terms of required number of $\xxgate$ gates. For the FT regime, the resource cost is measured in $\tgate$ gates, where each FT $\tgate$ gate requires tens of hardware-implementation-level $\xxgate$ gates \cite{ar:BH,ar:Campbell}. Shown are the water-molecule ground-state computation (Water)\cite{ar:VQE}, Heisenberg-Hamiltonian simulation (Heisenberg)\cite{ar:HINT}, maximum-cut optimization (QAOA)\cite{ar:QAOA}, the quantum Fourier transform (QFT)\cite{ar:QFT}, integer factoring (Shor)\cite{ar:Kutin}, generative-model quantum machine learning (QML)\cite{ar:GenMod}, Jellium- and Hubbard-model simulations (Jellium, Hubbard)\cite{ar:Babbush}, and the Femoco simulation (Femoco)\cite{ar:NF}. Grover's algorithm \cite{ar:Grover} implementation (not shown) that solves known difficult satisfiability problems \cite{ar:kSAT} requires ${\gtrsim} 2000$ qubits and ${\gtrsim} 2{\cdot}10^{27}$ $\tgate$ gates. See Supplementary Information (SI) section \ref{RR} for details.
 }
\label{fig:gatecounts}
\end{figure}

Highlighting the importance of efficient and robust implementation of $\xxgate$ gates, \fig{gatecounts} shows the resource requirements for various quantum computations. For this figure and for near-term, pre- fault-tolerant (FT) quantum computers, we considered variational quantum eigensolvers that compute the ground state of the water molecule \cite{ar:VQE}, a material spin-dynamics undergoing state-evolution according to the Heisenberg Hamiltonian \cite{ar:HINT}, a quantum approximate optimization algorithm addressing a maximum-cut problem relevant for various optimization problems \cite{ar:QAOA}, the widely-employed quantum Fourier transform subroutine \cite{ar:QFT}, quantum factoring of a 1024-bit integer \cite{ar:Kutin}, which is meaningful for cybersecurity, and data-driven quantum-circuit learning for certain visual patterns \cite{ar:GenMod}. The resource-cost metric used in the pre-FT regime considered here is the gate count for two-qubit $\xxgate$ gates, since these are the gates that limit algorithm performance. For the FT regime, in addition to the FT-regime-optimized versions of Heisenberg-Hamiltonian simulations, the quantum Fourier transform, and integer factoring, we considered Jellium- and Hubbard-model simulations for condensed-matter systems \cite{ar:Babbush}, the Femoco simulation \cite{ar:NF}, relevant for a certain nitrogen-fixation process that can help make fertilizer production more efficient, and solving difficult instances of satisfiability problems \cite{ar:Grover,ar:kSAT}. The resource-cost metric used for the FT regime is the number of $\tgate$ gates. Note that each $\tgate$ gate in FT quantum computing requires, e.g., a distillation process, typically referred to as a magic-state factory \cite{ar:BH,ar:Campbell}. Each distillation process for the $\tgate$-gate implementation in the FT regime requires at least a few tens of two-qubit gates, such as $\xxgate$ gates, at the native, hardware-implementation-level \cite{ar:BH,ar:Campbell}. Optimizing the $\xxgate$-gate implementation on a TIQIP is therefore critical for both pre-FT and FT quantum computing, and the construction 
and experimental demonstration 
of robust, power-optimal pulses for $\xxgate$-gate implementation on a TIQIP is the focus of this paper.

\section{Power-optimal two-qubit entangling gate}
There is only so much power that 
optical components can withstand. 
But more importantly, 
increased power leads to 
reduced gate fidelity due to, e.g.,  
carrier coupling \cite{WEIZ-PRL}, 
ion-ion crosstalk \cite{UMD,EASE}, 
and spontaneous emission from 
intermediate Raman levels \cite{GREENPAP}. 
Therefore, 
it is important to construct 
gates that, 
in addition 
to being stabilized against 
control-parameter fluctuations, 
require the least amount 
of power possible, i.e., they need 
to be power-optimal. In this paper 
we present a comprehensive, scalable approach 
to the construction of stabilized, 
power-optimal $\xxgate$ gates that is based on 
the M{\o}lmer-S{\o}rensen Hamiltonian of an ion chain 
interacting with 
laser pulses: 
\begin{equation}
\label{eq:Hamilton}
    H_{MS} = \sum_i \sum_p \eta_p^i g_i(t) 
    [ a_p\exp(-i\omega_p t)\sigma_x^i] + h.c.  
\end{equation}
Here $i$ and $p$ 
label the ions and the motional modes, 
respectively, $\eta_p^i$ is the Lamb-Dicke parameter,
$\sigma_x^i$ and $g_i(t)$ are the Pauli-$x$ operator 
and the pulse function   
acting on ion $i$, 
and $\omega_p$ and $a_p$ are the frequency 
and mode operator of 
motional mode $p$, respectively. 
A judicious choice of 
pulse functions generates an
$\xxgate$ gate that
induces entanglement between two 
trapped-ion qubits, defined by the unitary operator
\begin{equation}
\xxgate(\theta_{ij}) = e^{-i\theta_{ij}(\sigma_x^i
\sigma_x^j)/2},
\label{eq:XXUnitary}
\end{equation}
where $\theta_{ij} = 
4\chi_{ij}$ denotes the degree of entanglement between ions $i$ and $j$.
To induce the desired $\xxgate$ gate in practice, 
all motional modes of the ion 
chain need to be decoupled from the computational 
states of the qubits at 
the end of the gate operation
\cite{MS-1,IonQ,UMD}, 
leaving only the spins entangled. 
For an $N$-qubit system, 
and assuming that the same pulses are 
directed at ions $i$ and $j$, 
these constraints are of the form
\small
\begin{equation}
\label{eq:alpha}
\alpha_p = 
\int_0^\tau g(t) e^{i\omega_p t} dt = 0
\rightarrow
\sum_{n=1}^{N_A} M_{pn} A_{n} = 0,
\end{equation}
\normalsize
where $\tau$,  
a free parameter, is the pulse duration. 
The degree of entanglement between qubits $i$ and $j$ is obtained as
\small
\begin{align}
\label{eq:chi}
\chi_{ij} &= \sum_{p=1}^N \eta_p^i \eta_p^j \int_0^\tau dt_2 
\int_0^{t_2} dt_1 g(t_2) g(t_1) \sin[\omega_p(t_2-t_1)]
\nonumber \\
\rightarrow
\chi_{ij} &= \vec{A}\,^T D \vec{A} = \vec{A}\,^T S \vec{A}.
\end{align}
\normalsize

To find a power-optimal pulse, we require that the norm of $g$ is minimized, while $g$ still satisfies \eq{alpha} and \eq{chi} exactly.  
This 
can be achieved by
expanding $g$ in a complete basis,  
e.g., the Fourier-sine 
basis 
according to  
$g(t)=\sum_{n=1}A_n\sin(2\pi n t/\tau)$,
which spans the entire 
desired function space over 
the gate-time interval $\tau$. 
Restricted to a finite sub-space with basis-function amplitudes $A_n$, $n=1,2,..,N_A$, with sufficiently many $N_A$ basis functions, the constraint \eq{alpha} can be written in linear algebraic form as shown on the right-hand side of \eq{alpha}, where $M_{pn}$ is the time integral of the product between the $n$th basis function and $e^{i\omega_p t}$. Therefore, to satisfy \eq{alpha}, all that is required is to draw amplitudes $A_n$ from the null-space of $M$ \cite{GR}, where the null space is defined as the vector space that is mapped to zero under the action of $M$. Similarly, the constraints \eq{chi} can be denoted in linear algebraic form, as stated in the second line of \eq{chi}, where the matrix $D$ has elements $D_{nm}$, defined as the $p$-sum of the double integrals in \eq{chi} of the product between $\sin[\omega_p(t_2-t_1)]$ and the $n$th and $m$th basis functions that stem from expanding the two $g$ functions. Thus, defining the symmetric matrix 
$S = (D+D^T)/2$, 
\eq{chi} can be satisfied, including the requirement of minimal norm of $g$, by finding the appropriate linear combination of the null-space vectors of $M$ that combine to the eigenvector of $R$ with maximal absolute eigenvalue, where $R$ is the null-space projected 
matrix $S$. 

Our approach is linear and satisfies the two constraints \eq{alpha} and \eq{chi} exactly.
Since \eq{alpha} 
can be split into 
even and odd 
symmetry components,
it is possible to consider only $N$ out of $2N$ real constraints of \eq{alpha} and induce, at our discretion, a pulse that is symmetric or anti-symmetric about its center. Additionally, because, e.g., the Fourier basis is complete in its respective symmetry class, the resulting $g(t)$ is provably optimal in minimizing the norm of $g(t)$, which corresponds to 
minimizing 
the average power required to induce a $\xxgate$ gate. There is no iteration of any kind necessary. For instance, searching for an optimal solution in the parameter space, such as in \cite{AM,AM2,FM,PM,PARA1,PARA2,EASE}, is not necessary 
in our approach. 
Since only matrix 
operations are required 
to arrive at the optimal 
pulse solution, the 
optimal pulse is 
obtained in time $O(N_A^3)$ 
\cite{NUM-REC}. 

\begin{figure*}
\includegraphics[width=\textwidth]{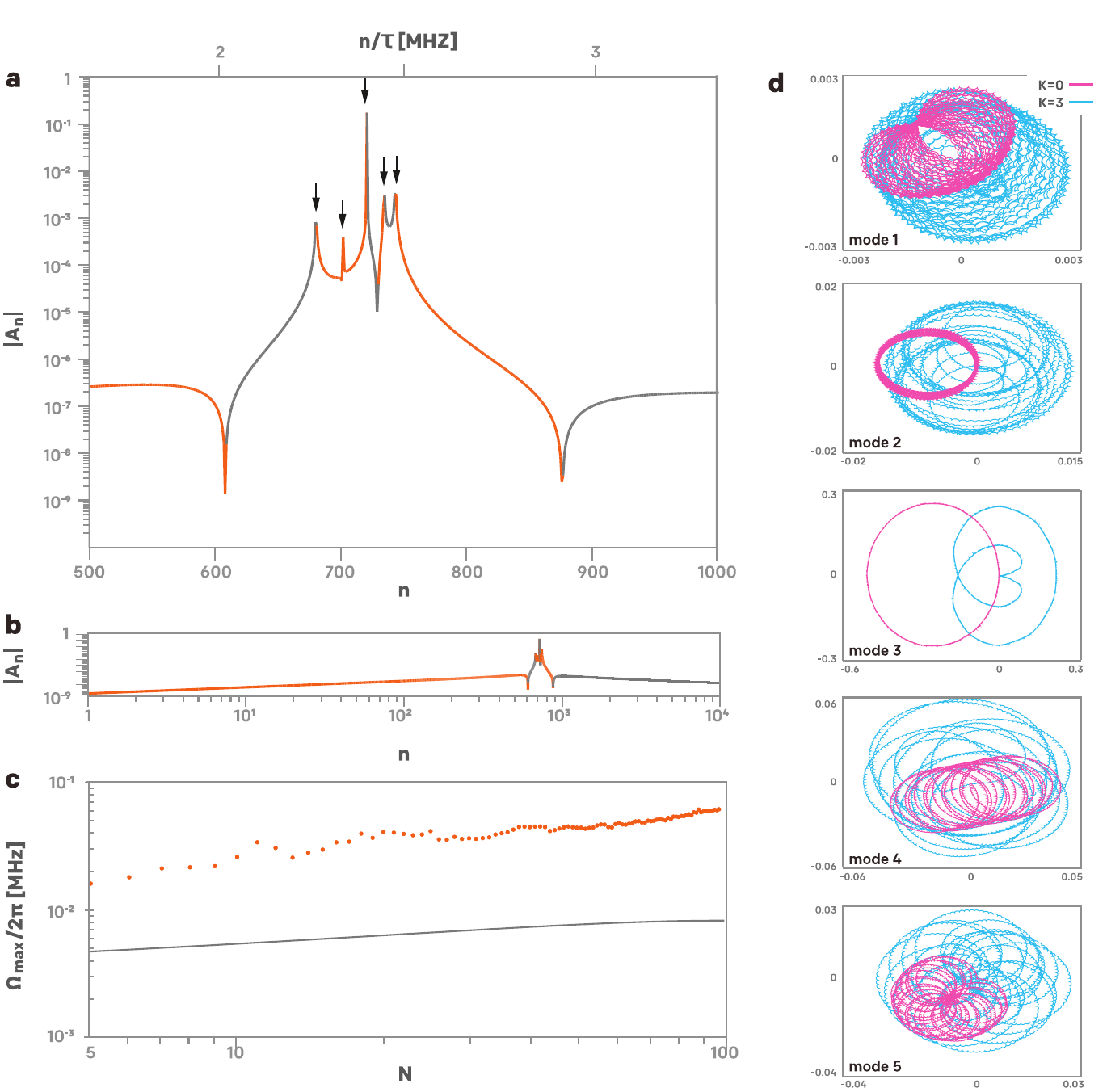}
\caption{Power-optimal pulse function $g$ and spin-dependent force applied to an ion qubit. 
See SI section \ref{PG} for the relevant parameters of the sample 
case of five qubits considered here. 
{\bf a}.
Fourier-sine coefficients $|A_n|$ of the pulse function $g(t) = \sum_n A_n \sin(2\pi nt/\tau)$, $\tau=300\mu$s. The tails of $|A_n|$ decay according to $\sim 1/n$. 
The arrows indicate the locations 
of the mode frequencies.
The signs of $A_n$ are color coded, i.e., negative $A_n$ are depicted with orange and positive $A_n$ are depicted with gray. The top scale shows the basis frequencies $n/\tau$, which 
are in resonance with the mode frequencies at the locations of the arrows.
{\bf b}. Same as {\bf a} but for a basis of 10,000 states. The main features 
occur at the same positions in $n$ and in frequency, which shows convergence 
and basis independence. 
{\bf c}. Scaling of the maximal power, $\max_t g(t)$ as a function of the number of qubits $N$. Gate time $\tau=500\mu$s. Orange circles: numerical results, heuristic bound. Gray curve: analytic bound derived in SI section \ref{APPCS}. {\bf d}. The time-dependent displacement in the phase space of ion number 1, where $K=0$ (pink) and $K=3$ (blue) are shown. 
The trajectories start and end at the origin.}
\label{fig:gfunction}
\end{figure*}

\Figs{gfunction}{\bf a} and {\bf b} show the amplitudes 
$A_n$ for a sample pulse function of the form 
$g(t) = \sum_n^{N_A} A_n 
\sin(2\pi nt/\tau)$ 
for $N_A=10000$ and $\tau=300\mu$s. 
As expected, the $|A_n|$ are large for $2\pi n/\tau \approx \omega_p$, 
showing that, to induce the desired $\xxgate$ interaction 
between two qubits via motional modes, the frequency 
components of the pulse function $g(t)$ need to be 
reasonably close to the motional-mode frequencies. 
We confirmed that a $N_A=1000$ basis-function solution 
essentially results in the same $A_n$ spectrum, visually 
indistinguishable from that with $N_A=10,000$, when overplotted.
This demonstrates the robustness of our method with respect to the basis size.  
 
\begin{figure}
\includegraphics[scale=0.35,angle=0]{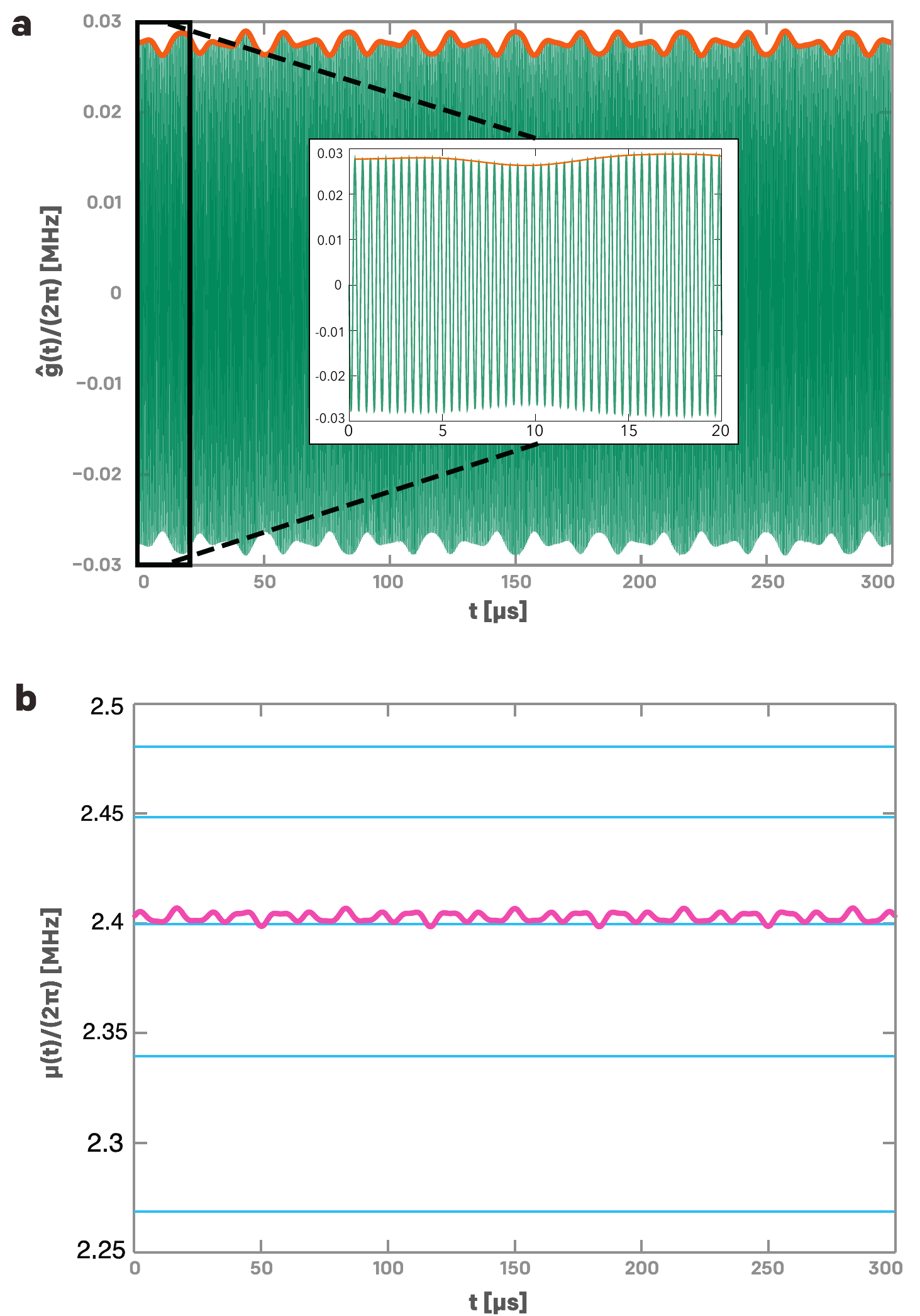}
\caption{Optimal $\tau= 300\,\mu$s pulse function 
for a maximally entangling $\xxgate$ gate between ions $i=1$ and 
$j=3$, for five ions with $N_A=1000$ basis states. 
The expansion coefficients 
$A_n$ of this pulse are shown in Fig.~\ref{fig:gfunction}{\bf a}. 
{\bf a}. The optimal pulse function $\hat g(t)$ (thin green line) 
with its amplitude function (thick orange line), 
obtained by demodulating 
$\hat g(t)$ 
as detailed in SI section \ref{denom}.
{\bf b}. Detuning function $\mu(t)$ obtained by 
frequency demodulating the pulse function, 
using the method described in SI section \ref{denom}.
The frequencies of the motional modes are  
shown as the five horizontal lines. 
The sample motional-mode frequencies used 
to generate this pulse 
are listed in Table~\ref{Tab1} and the set 
of $\eta$ parameters used are listed in Table~\ref{Tab2}
in SI section \ref{PG}.
}
\label{fig:AF}
\end{figure}

The pulse function $g(t)$ corresponding to the amplitudes $A_n$ 
shown in \Figs{gfunction}{\bf a} and {\bf b} is shown 
as the green line in Fig.~\ref{fig:AF}(a). Since $g(t)$ 
is a fast-oscillating function, it is instructive to 
write it in the form 
\begin{equation}
    g(t) = \Omega(t) \sin[\psi(t)], 
\label{g-decomposed} 
\end{equation}
where $\Omega(t)$ is the envelope function of the pulse 
(orange line in Fig.~\ref{fig:AF}{\bf a} and 
$\psi(t) = \int_0^t \mu(t')dt'$ is the phase function, 
where $\mu(t)$ is the 
detuning function 
\cite{IonQ,UMD}. 
We see that the amplitude of the pulse function is relatively flat, 
which implies that 
the average power minimization obtained by 
the $g$-norm minimization is essentially as good as 
minimizing the peak power of the pulse.
Figure~\ref{fig:AF}{\bf b} shows the detuning function $\mu(t)$. 
Consistent with the large Fourier amplitudes near the motional-mode 
frequencies in \fig{gfunction}{\bf a}, the demodulated frequency 
hovers around the motional-mode frequencies.

Because the minimal-power pulse function can be determined efficiently, 
it is straightforward to investigate the power requirement 
of the optimal pulse as a function of system size. \Fig{gfunction}{\bf c} 
shows the maximal power of the optimal pulse $\max_t g(t)$, 
obtained with $N_A = 1000$, for system sizes ranging from 5 to 100 ions. 
The power is consistent with our analytical bounds (see SI section \ref{APPCS}). 
Additionally, since according to the analytical results the power 
requirement is inversely proportional to the gate duration, 
the power optimum implies gate-time optimum for a given power budget. 
Thus, for a given amount of maximally available power, 
the power-optimal pulse is the fastest possible for $\xxgate$ gate execution. 
The ion displacement in position-momentum phase space for 
each mode $\omega_p$ entering into the 
computation of our 
sample pulse function $g(t)$ shown in \figs{gfunction}{\bf a} and {\bf b}, 
is shown in \fig{gfunction}{\bf d}.

\section{Control-pulse stabilization}
\label{CPS} 

Because the pulse function is constructed using a completely linear method, any additional linear constraints may be added, which still results in a power-optimal pulse when generated according to the steps discussed in the previous section. 
As an example, we show here how 
to stabilize the pulse against errors in external parameters, such as 
mode-frequency fluctuations. 

To stabilize against fluctuations of $\omega_p$, we start by expanding the number of constraints \eq{alpha}. Explicitly, we add 
\small
\begin{equation}
\label{eq:moments}
\frac{\partial^k}{\partial \omega_p^k} \int_0^\tau g(t) e^{i\omega_p t} dt = 0
\rightarrow
\sum_{n=1}^{N_A} M^{(k)}_{pn} A_{n} = 0,
\end{equation}
\normalsize
where $k$ denotes the order of stabilization. 
Since the additional constraints in 
\eq{moments} are linear, all we need to do to stabilize the pulse up to $K$th order is to include the additional linear equations \eq{moments} in the coefficients matrix $M$. The decoupling between the computational states of the qubits and the motional modes is thus achieved exactly, and the pulse is stabilized by 
using up $N(K+1)$ degrees of freedom.

\begin{figure*}
\centering
\includegraphics[width=\textwidth]{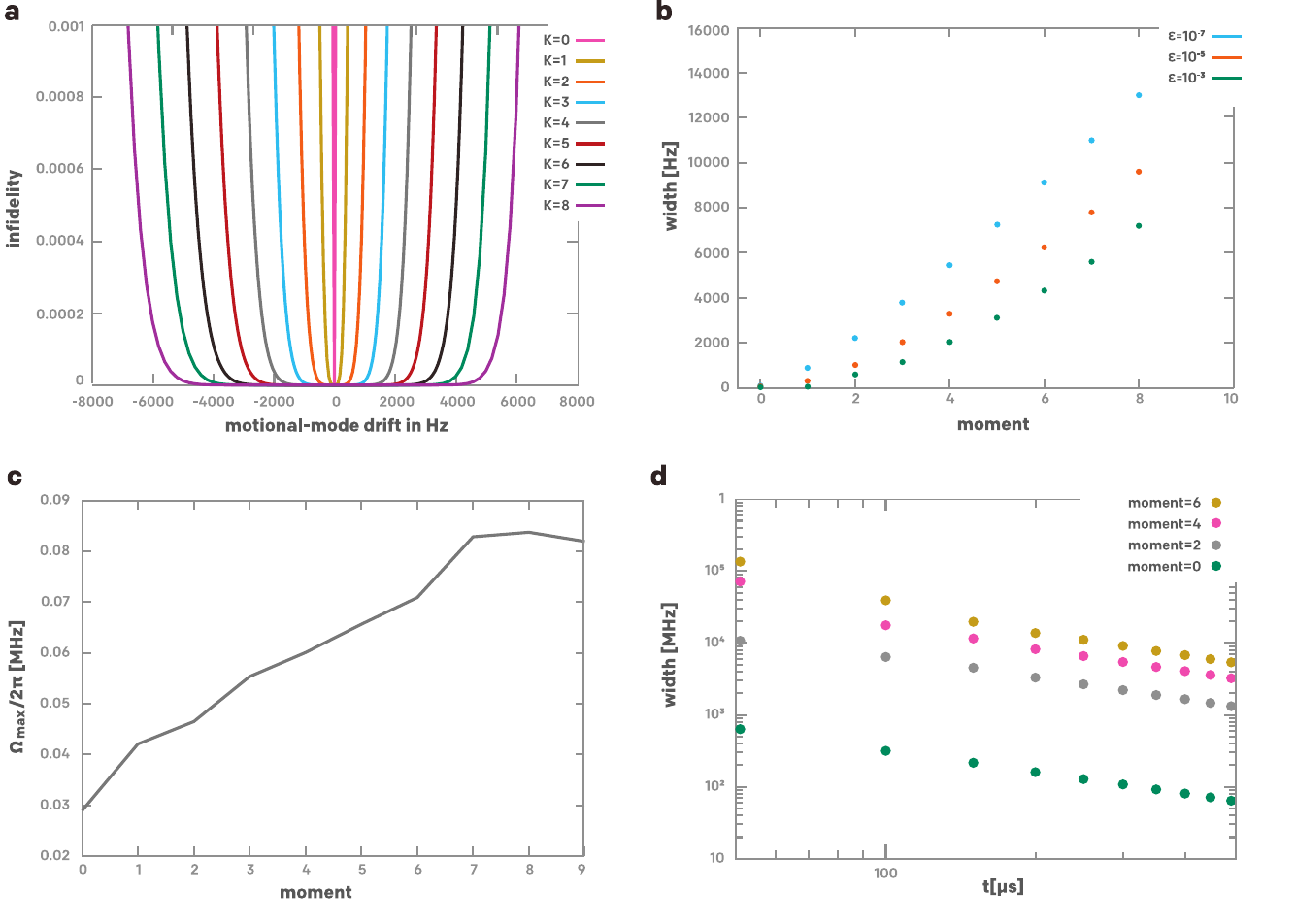}
\caption{Stabilization of the control pulses. {\bf a}. The infidelity (see SI section \ref{Stab} for detail) as a function of the motional-mode frequency drift $\Delta f$. All mode frequencies were drifted according to $\omega_p \mapsto \omega_p + 2\pi \Delta f$. {\bf b}. The width of the infidelity curves in {\bf a} for various error tolerances $\epsilon = 10^{-3}, 10^{-5}, \text{ and } 10^{-7}$, as a function of the highest moment $K$ of stabilization. {\bf c}. The maximal power requirement $\max_t|g_K(t)|$ of the control pulses, as a function of the order or moment of stabilization $K$. The power requirement suggests a linear scaling in the moment of stabilization. {\bf d}. The width of the infidelity curves for various different orders of stabilization $K=0, 2, 4, \text{ and } 6$, as a function of the gate duration $\tau$ for a fixed error tolerance level $\epsilon = 10^{-3}$. The data suggests $\sim 1/\tau$ scaling of the width.}
\label{fig:stabilize}
\end{figure*}

\Fig{gfunction}{\bf d} shows the phase-space trajectories for the stabilized pulse with $K=3$. Compared to the $K=0$ case, the general structure of the pulse with $K=3$ remains the same -- the frequency components are centered around the motional-mode frequencies and the phase-space closure is guaranteed. In \fig{stabilize}{\bf a}, the infidelity of stabilized 
pulses $K=0,1,..,8$ 
is shown as a function 
of the extent of the $\omega_p$ fluctuations. Considered are pulses with duration $\tau = 300\mu$s over the five-ion chain considered in the previous section. The widths of the infidelity curves, extracted at infidelity of 0.001, increase from ${\sim}0.1$kHz to ${\sim}13$kHz as $K$ is increased from 0 to 8 (see \fig{stabilize}{\bf b}). The power requirement of the stabilized pulses for each $K$ is shown in \fig{stabilize}{\bf c}; the requirement scales linearly in $K$. \Fig{stabilize}{\bf d} shows the width-scaling for each $K$ as a function of different choices of gate duration $\tau$. The effect of the stabilization increases inversely proportional to the gate duration.
 
The most straightforward 
way of experimentally 
implementing our AMFM pulses is via 
an arbitrary waveform 
generator (AWG) 
\cite{OXFORD-GATE,UMDPRL}. 
However, 
if an AWG is not available, implementation 
via the decomposition (\ref{g-decomposed}) 
(demodulation) 
is also possible 
(see Supplementary 
Information, section \ref{denom},
for more detail).

In various 
contexts our protocol has 
already been implemented, 
tested, and 
verified experimentally.  
In \cite{IonQ} it was used 
in its simpler 
amplitude-modulated 
version to benchmark 
one of the
IonQ quantum computers. 
In \cite{UMDPRL} it was 
used as the basis for 
demonstrating a new 
fidelity trade-off scheme 
called extended null-space 
(ENS) \cite{UMDPRL}. 
By sacrificing negligible 
amounts of fidelity, via ENS, 
an ``add-on'' to our basic 
AMFM scheme, additional 
power savings of up to an 
order of magnitude can 
be realized. 
This demonstrates 
that our protocol is 
extensible and 
adaptable. 

In both applications our 
power-optimal protocol 
has proved its 
experimental 
utility. 
 
While our protocol has already 
found experimental applications, 
its stabilizing effect 
in its exact 
[see (\ref{eq:alpha})] 
version, 
employing simultaneous 
amplitude and frequency 
modulation, has not yet 
been demonstrated. 
This is done in the following 
section. 

\section{Experimental Demonstration}
We implement the power-optimal, stabilized entangling gate on a trapped-ion quantum 
computer that has been described elsewhere in detail~\cite{IonQ}. A chain of seven ions is sideband cooled according to the protocol detailed in \cite{Sideband}. The two end ions are used to obtain equal spacing between the middle five individually addressed ions, which comprise the qubit register, and single- and two-qubit gate operations are driven by a two-photon Raman transition at 355~nm. 
As described in 
\cite{IonQ} 
the coherence of 
our quantum computer 
has been completely 
characterized. 
In particular, 
the fidelity 
of two-qubit gates 
was determined to 
be $\approx 96$\%, 
measured via 
parity contrast 
and 
partial tomography 
as described in 
\cite{IonQ,Gaebler2016,
Ballance2016}.

The propagator of 
(\ref{eq:Hamilton}) 
can be written 
in the form 
\cite{AM} 
\begin{equation}
    U = V 
    \xxgate(\theta_{ij}), 
\label{eq:U}
\end{equation}
where 
\begin{equation}
    V=\exp\left\{ 
-i\sum_{ip}
\eta_p^i
(\alpha_p a_p^{\dagger} + 
 \alpha_p^* a_p) 
 \sigma_x^i 
    \right\}
\end{equation}
and 
$\xxgate(\theta_{ij})$
is defined in 
(\ref{eq:XXUnitary}). 
For $\alpha_p=0$, 
which is guaranteed 
exactly according to 
our protocol, an 
AMFM gate solution, 
computed from the measured motional-mode spectrum, 
implements the unitary in 
(\ref{eq:XXUnitary}) 
over two qubits 
with $\approx 96$\% 
fidelity. 
Since $\alpha_p=0$ 
corresponds to zero 
mode-frequency drift, 
the $\approx 4$\% 
loss in fidelity 
are not due to 
mode-frequency 
drift but are due to 
other processes, such as 
beam-steering errors, 
laser-power fluctuations, 
fluctuating ambient 
electric and magnetic 
fields, etc..
 
Starting in the intial 
state $|00\rangle$, 
the ideal gate operator 
(\ref{eq:XXUnitary}) 
strictly 
preserves the even-parity 
population 
$P_{\rm even}=
P_{00}+P_{11}$, 
where $P_{mn}$ is the population 
in $|mn\rangle$ 
after the 
gate pulse is over. 
When motional-mode frequencies 
start to drift, 
$\alpha_p$ becomes 
nonzero and the 
propagator $V$ in 
(\ref{eq:U}) is 
``switched on''. 
This 
causes population 
transfer 
into the odd-parity 
states 
$|01\rangle$ and 
$|10\rangle$, 
which, in turn, causes 
a reduction of 
$P_{\rm even}$. 
Thus, $P_{\rm even}$ 
is a sensitive probe 
of stabilization of 
an AMFM gate against 
mode-frequency drift. 
We strongly 
emphasize that 
mode frequency drift 
causes a reduction of 
fidelity that is 
completely 
independent of the other 
sources of infidelity 
mentioned above. 
Thus, if the mode frequency 
drift is compensated, 
i.e., $V$ is 
``switched off'' 
by constructing a steering 
pulse $g(t)$ that assures 
$\alpha_p\approx 0$ to 
high order, 
as accomplished 
by our protocol, 
the gate 
fidelity is unchanged 
from its value at 
zero mode frequency 
offset. 
This was verified 
experimentally in 
\cite{UMDPRL} in 
the way of 
spot-checks using 
contrast 
measurements for 
selected values 
of mode frequency drift. 

\begin{figure}
\includegraphics[scale=0.7,angle=0]{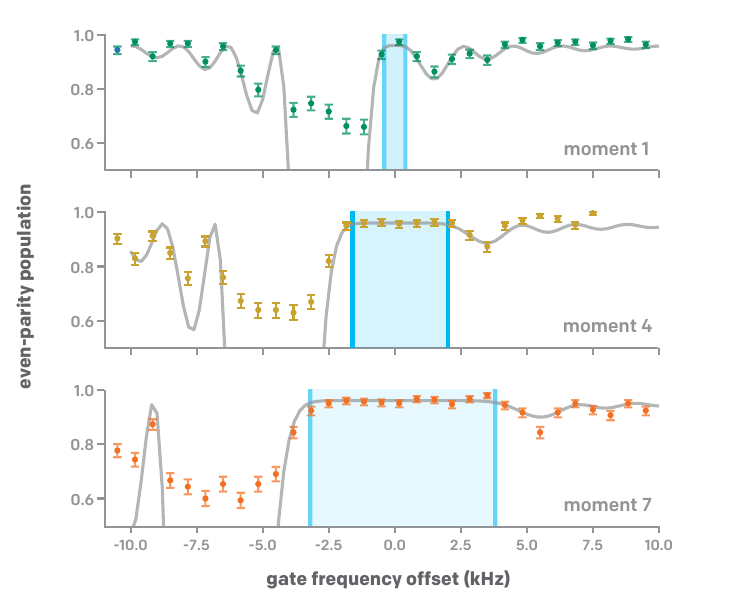}
\caption{Experimental demonstration of $K$-moment stabilized two-qubit gates. A single $\xxgate(\pi/2)$ gate is applied to the fiducial state $|00\rangle$, and the even-parity population is plotted as a function of the gate frequency offset. For increasing moment stabilization (shown are $K=1,4,7$), the region of high even-parity population increases, as indicated by the detuning-robust region shaded in blue. The gray line shows the analytical expression (\ref{MOM1}), valid in the low-error limit, with a 4\% offset to account for other sources of infidelity that are independent of the gate frequency. 
Notice that our experiment 
reproduces the oscillating
fine structure of 
the fidelity expression 
(\ref{MOM1}). 
Error bars on the experimental data are 1$\sigma$ confidence intervals, sampled from a binomial distribution, and each point represents 300 realizations of the experiment.
}
\label{fig:exp}
\end{figure}

Figure~\ref{fig:exp} shows
the
even-parity population $P_{\text{even}}$ 
as a function of 
mode-frequency 
offset, 
measured after 
applying an 
$\xxgate(\pi/2)$ 
gate
designed for zero 
mode-frequency 
offset, 
to the initially prepared two-qubit state 
$|00\rangle$. 
Three different 
pulses were used 
to implement $\xxgate(\pi/2)$, 
with 
moment 
stabilization orders $K=1,4,7$.
The  
grey
line in 
each 
frame of 
Fig.~\ref{fig:exp}
shows the analytical infidelity 
$(4/5)\sum_p \big[ (\eta_p^i)^2 + (\eta_p^j)^2 \big] |\alpha_p|^2$ 
\cite{GREENPAP}, 
with a constant 
4\% offset to account for 
the
other errors 
that are
independent of the gate-frequency offset.
The shaded region indicates the range of 
mode-frequency 
offsets 
for each moment 
stabilization where the theoretical 
fidelity $F > 0.99$. 
The width of the shaded 
region
can be seen 
to increase 
for increasing moment. 
In fact, for 
$K=1$ stabilization 
(top frame of 
Fig.~\ref{fig:exp})
acceptable fidelity 
is achieved only 
over a range of 
less than $1\,$kHz, 
outside of which 
$P_{\rm even}$ drops 
precipitously to around 
65\%. 
$K=4$ stabilization 
(middle frame of 
Fig.~\ref{fig:exp}) 
achieves stability 
over a mode-frequency 
range of about 
$\pm 2\,$kHz, 
a factor-2 improvement 
of allowed mode-frequency 
fluctuations accompanied 
by a restoration of 
the gate fidelity back up 
to $\approx 96$\%. 
In the interval between 
$1\,$kHz and $2\,$kHz 
we see an improvement 
of $P_{\rm even}$ of 
about 30\%. This is 
not a small effect. 
This is a sizable 
stabilization effect 
adding to the proven 
utility 
of our protocol. 
$K=7$ stabilization 
(bottom frame of 
Fig.~\ref{fig:exp}) 
shows a similar 
$\approx 30$\% 
improvement of 
$P_{\rm even}$ 
over an even larger 
frequency interval 
$\approx \pm 3\,$kHz), 
a factor-3 improvement 
of the stable frequency 
range. 
More details 
concerning 
our experiment
can be found in SI section~\ref{ImpDet}.

\section{Discussion}
\label{DISC} 

Application of our 
protocol is not limited to the 
case of the trapped ion-chain 
quantum computer architecture 
\cite{IonQ,UMD}. It can   
be applied to all quantum 
computing architectures that 
rely on phase-space closure, 
such as the QCCD architecture 
\cite{HONEYWELL}. 
It is even applicable to 
the superconducting quantum 
computer architecture that 
also relies on pulse shaping 
\cite{PHYS-TODAY}. 

Our work is most closely related 
to \cite{SL-PRL,WEIZ-PRL}, who 
use a multi-tone approach to 
construct stabilized two-qubit 
gates.   
However, there are substantial 
differences. For instance: 
(i) The authors of 
\cite{SL-PRL,WEIZ-PRL} 
formulate 
and demonstrate their protocol 
for the special case of 
trapped two-ion crystals, and 
since it relies 
on an analytical 
solution for the gate pulses, 
it is not scalable to $N$-ion, 
many-mode systems. 
(ii) The multi-tone approach 
in \cite{SL-PRL,WEIZ-PRL} requires 
the different tones (which can 
be interpreted as Fourier 
components) of $g(t)$ to be 
equi-spaced. While this is very 
closely related to our expansion 
of $g(t)$ in a Fourier-sine 
series, equi-spacing in the 
frequency 
of basis functions 
is not necessary 
in our approach. Any basis 
can be used to express $g(t)$.  
For instance, we found that short pulses
are better represented in a 
polynomial basis, whose 
Fourier components are not 
equi-spaced. Thus, any kind 
of chirped pulse 
can be used in our method. 
(iii) While, indeed, stability 
against several types of errors 
was demonstrated in 
\cite{SL-PRL,WEIZ-PRL}, these 
conditions, to much higher 
order than demonstrated in 
\cite{SL-PRL,WEIZ-PRL}, are 
more easily implemented 
according to our protocol, 
which requires only added 
linear equations for each 
stability condition or constraint. 
(iv) As discussed in SI 
sections~\ref{sec:Proj}, 
\ref{SOM-ASMM}, and 
\ref{SOM-ASHM}, our protocol 
can be extended to include 
stability against drifts 
in the degree of entanglement. 
(v) Power optimization was not 
considered in 
\cite{SL-PRL,WEIZ-PRL}. 
Thus our protocol, presenting 
a unified, scalable approach 
to arbitrarily stabilized $N$-ion gate 
construction,
goes far beyond 
the methods presented in 
\cite{SL-PRL,WEIZ-PRL} and 
is thus a substantial 
advance over existing methods. 
 
The ability to symmetrize the pulse solution gives rise to potential additional room for robustness with respect to errors. Since, e.g., the inner products between a symmetric pulse function $g(t)$ and the antisymmetric part of the constraint $e^{i\omega_p t}$ in \eq{alpha} are zero (see SI section \ref{SC} for details), akin to echos, as long as the pulse function is modified symmetrically due, e.g., to implementation defects, half of the null-space conditions in \eq{alpha} is still exactly satisfied, while leaving the error entirely in the part of the constraint with the opposite symmetry. The knowledge that the error lies in the oppositely symmetric part leaves room for a secondary echo, wherein the sign of the errors may be flipped. In case the errors cannot be manipulated to be echoed out in this particular symmetry, other symmetries may be considered at the pulse-construction level, 
rendering our approach an integrated protocol that can be designed to be robust against different symmetry classes of errors.

The methodology and paradigm used to construct our power-optimal pulses is general and can readily accept incomplete bases to result in pulses that are subjected to additional constraints.
Beyond the AM, FM, PM pulses that can be obtained by appropriately demodulating the pulse function 
according to (\ref{g-decomposed}),
a step-pulse approach that has been used in the literature \cite{AM,AM2,EASE} can also be derived. In fact, as shown in SI section \ref{STP}, by carefully tuning the gate duration time, even the beneficial symmetric pulse structure can be preserved. 

Adding to the generality is the application to the Efficient Arbitrary Simultaneously Entangling (EASE) gates \cite{EASE}, where any combination of quadratically many pairs of qubits can be entangled to any degree of entanglement. As detailed in SI section \ref{sec:EASE}, because our approach is linear 
and the EASE-gate approach is amenable to any linear approach, it is straightforward to adapt the pulse-construction method presented here to the EASE protocol. Together with the power- or 
time-optimality guaranteed in our pulse construction by design, then, the EASE gate equipped with our method enables one of the fastest ways to implement as many entangling gates as possible in a TIQIP.

The moment-stabilization adds robustness against those errors induced by not properly decoupling the computational states from the motional modes. However, unitary errors in the computational space may still linger, since the entanglement degree is sensitive to, e.g., $\omega_p$ fluctuations (see SI section \ref{sec:Sensitivity}). 
In practice, this may be fended off by calibrations, i.e., by monitoring how the degree of entanglement changes over time and adjusting the amplitude of the illuminating beams to compensate for this change. Note that the shape of the pulse does not change; amplitude scaling suffices. If frequent calibrations are impractical, active stabilization in the 
entanglement degree $\chi_{ij}$ may be 
implemented, which involves finding pulses 
that, apart from satisfying (\ref{eq:alpha}), 
(\ref{eq:chi}), and (\ref{eq:moments}), 
also satisfy 
\begin{equation}
\label{eq:chistab}
    \frac{\partial^k\chi_{ij}}
    {\partial\omega_p^k} = 0,\ \ \ 
    k=1,2,\ldots,K_{\chi}, 
\end{equation}
where $K_{\chi}$ is the maximal 
order of desired $\chi$ stabilization. 
In SI sections \ref{sec:Proj}, \ref{SOM-ASMM}, and \ref{SOM-ASHM}, 
we offer three methods 
which either 
achieve (\ref{eq:chistab}) approximately 
(see \ref{sec:Proj}) or completely (see \ref{SOM-ASMM} and \ref{SOM-ASHM}). 
While the projection method presented 
in SI section \ref{sec:Proj} achieves 
(\ref{eq:chistab}) only approximately, 
and only for $K_{\chi}=1$, 
it does reduce the 
errors originating from fluctuating 
mode frequencies substantially. 
A qualitative improvement over the 
projection method are the 
two-pulse moments method
presented 
in \ref{SOM-ASMM} and  
the hybrid method, presented in \ref{SOM-ASHM}. 
These two methods fulfill 
(\ref{eq:chistab}) exactly, and 
consequently actively stabilize 
the degree of entanglement over 
a large interval of frequency drift 
[see SI section \ref{SOM-ASMM}, Fig.~\ref{fig:chi-stabilize}{\bf a} 
and SI section \ref{SOM-ASHM}, Fig.~\ref{fighybrid}]. 
However, 
constructing the two pulses characterizing 
these methods is 
computationally more expensive and 
the pulses require more power.
Therefore, 
whether to use the simple projection method 
described in \ref{sec:Proj} or 
the two-pulse methods described 
in \ref{SOM-ASMM} and \ref{SOM-ASHM}, depends on experimental conditions 
and available computational resources. 

Stabilization against other parameters, such as the Lamb-Dicke parameters $\eta_p^i$ or the amplitude of the laser beam, encoded in the norm of $g(t)$, is also possible. Notice that $\eta_p^i {\mapsto} \eta_p^i (1{+}\Delta_{\eta_p^i})$ or $g(t) {\mapsto} g(t)(1{+}\Delta_g)$, where $\Delta_{\eta_p^i}$ and $\Delta_g$ are small constants, does not affect the decoupling condition \eq{alpha}. The effect of the errors is confined to the degree of entanglement \eq{chi}, i.e., $\chi_{ij} {\mapsto} \chi_{ij} {+} \Delta\chi_{ij}$, where $\Delta\chi_{ij}$ is the error in $\chi_{ij}$ that arises from $\Delta_{\eta_p^i}$ and $\Delta_g$. As discussed in SI section \ref{sec:Broad}, this can be adequately compensated by, e.g., a broadband compensation sequence applicable for the two-qubit case.

We note that, while technically challenging, in principle, it is possible to directly implement our Fourier-basis pulse solution $g(t)$ using a multi-tone laser. As shown in SI section \ref{sec:Direct}, by implanting $N_A$ different colors with different amplitudes to the beams that address ions, then locking the phases of them, we can induce the desired evolution of the $\xxgate$ on a TIQIP. The technique here is similar to the discrete multi-tone, widely used in communication lines. The development of the technology in the optical regime remains as a 
promising avenue for future research. 

\section{Conclusion}
Formally speaking, there are infinitely many smooth solutions that qualify as adequate pulse functions. 
Out of these infinitely many possible solutions, our protocol extracts the power-optimal solution without any iterations or parameter scans. Including symmetry and stabilization, the solution is also robust against errors.
With an AWG, 
as already demonstrated 
\cite{UMDPRL}, 
our AMFM pulses can be 
implemented directly 
experimentally 
to produce an 
optimal 
$\xxgate$ gate.
If AWG technology is not 
available, as in our 
experiments reported in
this paper, 
implementation via 
demodulation (see 
SI section \ref{denom}) 
is also an option. 
Demodulation of our AMFM 
pulses also shows that their 
amplitude function is nearly 
flat, i.e., 
average power 
minimization is an excellent approximation of 
maximal power minimization. Indeed, an exact analytical bound on power and its comparison to the demodulation results shows that the optimal solution is close to the bound. With the $\xxgate$ gate implemented using the pulse constructed according to our protocol, just about any quantum algorithms can now be implemented with minimal power requirement, or in the shortest possible time for a given power budget, at the two-qubit gate, physical implementation level. This provides decisive advantages in improving both noisy, near-term TIQIPs and fault-tolerant TIQIPs to come in the future.

\section*{Acknowledgements}
The authors would like to acknowledge Coleman Collins for assistance with visual illustrations and Dr. Ming Li for helpful discussions.

\section*{Supplementary Information}

\renewcommand\theequation{S\arabic{equation}}    
\renewcommand\thetable{S\arabic{table}}    
\renewcommand\thefigure{S\arabic{figure}}    
\renewcommand\thesubsection{S\arabic{subsection}}
\renewcommand\thesection{}
\setcounter{subsection}{0}    
\setcounter{equation}{0}    
\setcounter{table}{0}    
\setcounter{figure}{0}    

\subsection{Resource requirement}
\label{RR}

In this section, we detail the methods used to obtain both pre- fault-tolerant (FT) and FT-regime resource requirements presented and illustrated in the main text, Fig.~1. The cases considered are (i) the water molecule ground-state energy estimation \cite{ar:VQE}, (ii) Heisenberg-Hamiltonian simulation \cite{ar:HINT}, (iii) quantum approximate optimization algorithm solving a maximum-cut problem \cite{ar:QAOA}, (iv) the quantum Fourier transform \cite{ar:QFT}, (v) integer factoring \cite{ar:Kutin}, (vi) data-driven quantum circuit learning \cite{ar:GenMod}, (vii) Jellium and Hubbard-model simulation \cite{ar:Babbush}, and (viii) the Femoco simulation \cite{ar:NF}. 

For case (i), we considered pre-FT HF+7 and HF+21 cases, where HF denotes the Hartree-Fock method detailed in \cite{ar:VQE}, and 7 and 21 denote different approximation qualities. The $\xxgate$ gate counts for the two cases are available in Fig.~2{\bf b} of \cite{ar:VQE}.

For case (ii), we considered the Heisenberg Hamiltonian applied to spins with their connectivity specified by $(k{-}d{-}n)$ graphs, where $k$ denotes the degree, $d$ denotes the distance, and $n$ denotes the number of vertices of the graph. Specifically, the graphs 
considered are $(3{-}5{-}70)$, $(4{-}4{-}98)$, and $(5{-}3{-}72)$. For the pre-FT cases we used $\cnotgate$ gate counts reported in the pre-FT part of Table~I of \cite{ar:HINT}. For the FT cases, we used $\tgate$-gate counts reported in the FT part, specifically the RUS part, of the same table.

For case (iii), we considered the quantum approximate optimization algorithm in the pre-FT regime with eight stages, based on its performance compared to the well-known instance of semidefinite programming called Goemans-Williamson approximation algorithm \cite{ar:GW}. The graphical representation of how the quantum algorithm solving the maximum cut problem performs with stage numbers $2^0,2^1,..,2^5$ may be found in Fig.~2 of \cite{ar:QAOA}. Each stage requires $n(n-1)/2$ $\xxgate$ gates, as can be seen from Eq.~(7) of \cite{ar:QAOA}.

For case (iv), we considered the approximate quantum Fourier transform \cite{ar:QFT}, where all controlled-rotation gates with rotation angles less than $\pi/2^b$, $b = \log_2(n)$, where $n$ is the number of qubits, are removed. For the pre-FT regime, one $\xxgate$ gate was expended per controlled-rotation gate. For the FT regime, see Table 1 of \cite{ar:QFT}.

For case (v), we used the implementation 
presented in \cite{ar:Kutin}. While an explicit resource cost is not available, an estimate is available in section A of the appendix of \cite{ar:HeisenNN}. The implementation in \cite{ar:Kutin} uses $4n^3 + O(n^2 \log(n))$ gates and $3n + 6 \log(n) + O(1)$ qubits, assuming an arbitrary two-qubit gate may be implemented. For the pre-FT regime, each arbitrary two-qubit gate costs three $\cnotgate$ or $\xxgate$ gates, as per \cite{ar:Markov}. For the FT regime, see the discussion section A of the appendix of \cite{ar:HeisenNN}, which results in $16 n^3$ $\tgate$ gates.

For case (vi), we largely base the resource counts on Table~1 of \cite{ar:GenMod}, where several sample instances of bars-and-stripes patterns are explicitly considered for $n$ ranging from 4 to 100. The expected $\xxgate$ gate counts are computed assuming the all-to-all connectivity available in the trapped-ion quantum information 
processor 
(TIQIP), and we used four layers in the training circuit (see Fig.~1 of \cite{ar:GenMod} for further information) that worked well for a small system with $n=4$.

For case (vii), Tables 3 and 4 of \cite{ar:Babbush} detail the FT resource-cost for several different cases. 
 
 For case (viii), see Table 1 of \cite{ar:NF} for the Femoco simulation. We used a serial version of structure 1 with accuracy of simulation of $10^{-3}$ Hartree. Note this case appears to demand a fairly large amount of resources compared to other examples considered herein. This is due in part to the other examples being tailor-developed and co-designed for minimal resource requirements. Subsequent development since \cite{ar:NF} has led to a reduction in the resource requirements by several orders of magnitude, inching closer to the cluster of points that appear in Fig.~1 of 
 the main text. See \cite{lee2020even} for details.

We also considered 
Grover's algorithm solving certain difficult instances of a Boolean satisfiability problem \cite{ar:Grover}
with $n$ variables and $m$ clauses. Specifically, we considered ``hole12'', ``Urq7\_5'', ``chnl11x20', and ``fpga13.12'' problems, where the names were taken verbatim from Table 1 of \cite{ar:kSAT}. To construct the FT circuit, we used $k$-control Toffoli gates to implement the Grover oracle \cite{ar:Grover}, where $k$ is the length of a clause. Specifically, we used $m$ clean ancilla qubits to compute the satisfiability of $m$ clauses individually, and used a $m$-control Toffoli gate with an additional ancilla qubit to implement the oracle. Whenever possible, we used relative Toffoli gates in \cite{ar:reltof} to reduce the $\tgate$ counts, while keeping track of the number of recyclable ancilla qubits in implementing the multi-control Toffoli gates. Together with a $n$-control Toffoli gate for the Grover diffusion operator, we obtain for ``hole12'' 2053 qubits and $2.094 \cdot 10^{27}$ $\tgate$ gates, for ``Urq7\_5'' 4627 qubits and $2.031 \cdot 10^{40}$ $\tgate$ gates, for ``chnl11x20'' 8879 qubits and $4.931 \cdot 10^{70}$ $\tgate$ gates, and for ``fpga13.12'' 2717 qubits and $1.538 \cdot 10^{39}$ $\tgate$ gates, where we used $\lceil \pi/4 \cdot \sqrt{2^n}\rceil$ iterations for near-optimal results. Of course it is challenging to realize 
on the order of $10^{27},\ldots,10^{70}$ 
quantum gates. 
However the presented scaling corroborates 
the need for efficient implementations of 
quantum gates in less demanding circumstances.

\subsection{Ising gate on a trapped-ion quantum information processor}
\label{Ising}
\begin{figure}
\includegraphics[scale=0.7,angle=0]{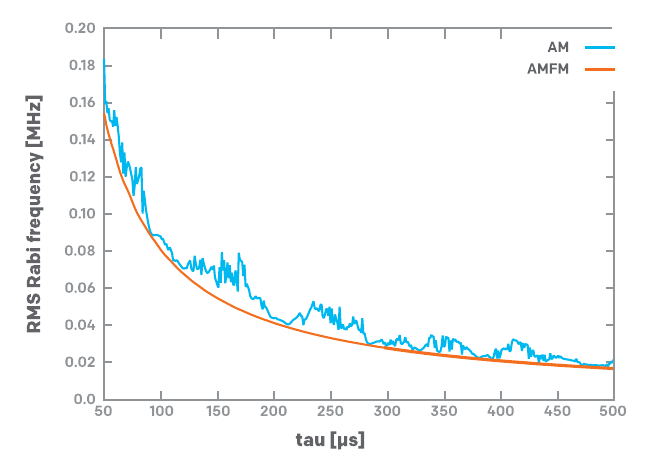}
\caption {
Comparison of RMS Rabi frequency required for fully entangling gates between AM and AMFM pulses. The power-optimal AM and AMFM pulses are computed in intervals of $1\mu$s. For every gate time, the detuning of AM pulses is scanned in steps of 10Hz, and the lowest RMS Rabi frequency obtained is presented on the plot. The gates are computed for qubits $i=1,j=3$ in a five-ion chain, without including any stability conditions.  
      }
\label{fig:amfm-vs-am}
\end{figure}

The participating 
ions of an Ising $\xxgate$ gate couple to all motional modes \cite{MS-1}, 
and have to be decoupled 
from the motional modes 
at the end of the gate. 
The relevant equations are 
\cite{AM}
\begin{align}
\label{eq:INTRO1} 
\alpha_{ip} &= \int_0^{\tau} \Omega_i(t) 
\sin[\Phi_i + \psi_i(t)] e^{i\omega_p t}\, dt = 0,
\nonumber\\
i&=1,\ldots,N, \ \ \ p=1,\ldots, P, 
\end{align}
where $\tau$ is the length of the pulse, 
$i$ is the ion number, 
$N$ is the total number of ions, 
$p$ is the mode number, $P$ is the total 
number of modes, 
$\Phi_i$ is the initial phase, 
$\omega_p$ are the motional-mode 
frequencies, and 
$\Omega_i(t)$ 
is the amplitude function, 
i.e., the time-dependent 
Rabi frequency.
The time-dependent phases $\psi_i(t)$ in 
\eq{INTRO1} are defined as 
\begin{equation}
\psi_i(t) = \int_0^t \mu_i(t')\, dt', 
\label{INTRO2}
\end{equation}
where  $\mu_i(t)$ is the detuning function. 
In order not to start the pulse abruptly, 
we require $\Phi_i=0$. For ease of presentation, 
we also assume, from now on, that the same 
pulse shape acts on all $N$ ions, such that, 
together with the assumption of vanishing initial phase, 
\eq{INTRO1} acquires the simplified form 
\begin{align}
\alpha_{ip} &= \int_0^{\tau} \Omega(t) 
\sin[\psi(t)] e^{i\omega_p t}\, dt = 0,  \nonumber \\
\ \ \ & i=1,\ldots,N, \ \ \ p=1,\ldots, P , 
\label{INTRO5} 
\end{align}
where 
\begin{equation}
\psi(t) = \int_0^t \mu(t')\, dt'. 
\label{INTRO22}
\end{equation}
If the pulse acts 
simultaneously on ions $i$ and $j$, 
the gate angle $\varphi_{ij}$ of the $\xxgate$ gate is given by 
\cite{AM}
\begin{equation}
\varphi_{ij} = \chi_{ij} + \chi_{ji} , 
\label{INTRO3}
\end{equation}
where 
\begin{align}
\label{eq:INTRO4}
\chi_{ij} &= \sum_{p=1}^P\, \eta_p^i \eta_p^j 
\int_0^{\tau} dt_2 \int_0^{t_2} dt_1
\Omega(t_2) \Omega(t_1) 
\nonumber\\
&\sin[\omega_p(t_2-t_1)] 
\sin[\psi(t_2)] \sin[\psi(t_1)] , 
\end{align}
and 
$\eta_p^i$ is the Lamb-Dicke parameter
\cite{LD}, which describes the coupling strength of ion 
number $i$ to motional-mode number $p$. 
A maximally entangling gate is 
achieved for $\varphi_{ij}=\pm\pi/4$. According to 
\eq{INTRO4}, $\chi_{ij}=\chi_{ji}$, i.e., 
$\varphi_{ij}=2\chi_{ij}$, so that 
a maximally entangling gate requires 
\begin{equation}
|\chi_{ij}| = \frac{\pi}{8}. 
\label{INTRO4a}
\end{equation}
Since both $\Omega(t)$ and $\sin[\psi(t)]$ are unknown, 
we combine them into one single pulse function 
\begin{equation}
g(t) = \Omega(t) \sin[\psi(t)]. 
\label{METH3}
\end{equation}
Thus, for given motional-mode frequencies 
$\omega_p$ and Lamb-Dicke parameters 
$\eta_p^i$, 
our task is to find a pulse $g(t)$, which 
solves (\ref{INTRO5}) and produces 
$|\chi_{ij}|=\pi/8$ with 
minimal power requirement. 
Known solution methods include amplitude-modulation 
techniques \cite{AM,AM2}, 
which require fixed detuning 
frequency $\mu_0$, frequency-modulation 
techniques \cite{FM}, 
which require a given 
shape of the pulse-envelope function 
$\Omega(t)$, and 
phase modulation \cite{PM}. 
Our approach goes beyond 
previously demonstrated approaches in that we modulate 
amplitude, frequency, and phase {\it simultaneously}. 
In addition, we use a {\it linear} method, which 
yields the optimal pulse shape directly, without 
any iterations or parameter searches, 
using exclusively linear-algebra 
techniques. 

Figure~\ref{fig:amfm-vs-am} 
shows a comparison between RMS Rabi 
frequencies,  
$[\langle \Omega^2(t)\rangle]^{1/2} / 
(2\pi)\,$ in MHz,  
required for fully 
entangling gates using AM and AMFM pulses. 
The data shows the lowest RMS Rabi 
frequencies obtained when scanning the gate time in steps of $1\mu$s. For every gate time, the detuning of AM pulses is scanned in steps of $10\,$Hz, and the lowest RMS Rabi frequency obtained is presented on the plot. The data points were computed without any added stabilization to external sources of errors. While the required RMS Rabi frequency is lower for AMFM pulses, the stability to external errors is comparable. The execution times of the classical computations required to find the power-optimal pulses for every gate time are comparable, with 0.73 seconds for an AMFM pulse and 0.93 seconds for an AM pulse. This, however, changes drastically 
if higher-order moment stabilization is 
included, which requires the AM pulse to 
be represented by a large number of 
segments. In this case, the number of 
segments will approach the number of 
AMFM basis states and the execution time 
of the AMFM method will be shorter 
by a large factor. 

\subsection{Symmetry classes}
\label{SC}

Since $g(t)$ is a real function, the $P$ complex 
equations (\ref{INTRO5}) for $p=1,\ldots,P$ 
are equivalent to 
$2P$ real equations 
\begin{align}
\int_0^{\tau} g(t) \cos(\omega_p t)\, &= 0, \ \ \ 
\int_0^{\tau} g(t) \sin(\omega_p t)\, = 0, \ \ \  \nonumber \\
&p=1,\ldots, P. 
\label{PSC1}
\end{align}
It follows that if (\ref{PSC1}) is satisfied, 
any linear combination 
\begin{equation}
h_p(t) = A_p\cos(\omega_p t) + B_p\sin(\omega_p t) 
\label{PSC2}
\end{equation}
satisfies 
\begin{equation}
\int_0^{\tau} g(t) h_p(t)\, = 0. 
\label{PSC3}
\end{equation}
We define two special linear combinations 
\begin{align}
h_p^{(+)}(t) &= 
\cos\left(\frac{\omega_p\tau}{2}\right)
\cos(\omega_p t) + 
\sin\left(\frac{\omega_p\tau}{2}\right)
\sin(\omega_p t) 
\nonumber\\
&= 
\cos\left[\omega_p \left(\frac{\tau}{2}-t\right)\right]  
\label{PSC4}
\end{align}
and 
\begin{align}
h_p^{(-)}(t) 
&= 
\sin\left(\frac{\omega_p\tau}{2}\right)
\cos(\omega_p t) -
\cos\left(\frac{\omega_p\tau}{2}\right)
\sin(\omega_p t) 
\nonumber\\
&=
\sin\left[\omega_p \left(\frac{\tau}{2}-t\right)\right],  
\label{PSC5}
\end{align}
which satisfy 
\begin{equation}
h_p^{(\pm)}\left(\frac{\tau}{2}-t\right) = 
\pm 
h_p^{(\pm)}\left(\frac{\tau}{2}+t\right), 
\label{PSC6}
\end{equation}
i.e., $h_p^{(+)}(t)$ and $h_p^{(-)}(t)$ are even and odd 
functions with respect to $\tau/2$. We also define 
\begin{equation}
g^{(\pm)}(t) = \frac{1}{2} 
\left[ g\left(\frac{\tau}{2}+t\right) \pm 
g\left(\frac{\tau}{2}-t\right) \right] , 
\label{PSC7}
\end{equation}
i.e., 
the even and odd components of the pulse $g(t)$. 
We call $g^{(+)}(t)$ the positive-parity pulse and 
$g^{(-)}(t)$ the negative-parity pulse. Then 
the $P$ equations 
\begin{equation}
\int_0^{\tau} g^{(\pm)}(t) h_p^{(\mp)}(t)\, dt = 0, 
\ \ \ p=1,\ldots, P 
\label{PSC8}
\end{equation}
are satisfied automatically, which implies that for 
given parity, we have to satisfy only $P$ real, 
nontrivial equations 
\begin{equation}
\int_0^{\tau} g^{(\pm)}(t) h_p^{(\pm)}(t)\, dt = 0, 
\ \ \ p=1,\ldots, P. 
\label{PSC9}
\end{equation}
In analogy to the definition of the two parities for 
the pulse function $g(t)$, we may also define 
even and odd pulse envelope functions, 
$\Omega^{(\pm)}(t)$, 
and even and odd detuning functions, 
$\mu^{(\pm)}(t)$, 
which are even and odd 
functions with respect to $\tau/2$ according to 
\begin{align}
\Omega^{(\pm)}\left(\frac{\tau}{2}-t\right) = 
\pm 
\Omega^{(\pm)}\left(\frac{\tau}{2}+t\right), 
\nonumber \\ 
\mu^{(\pm)}\left(\frac{\tau}{2}-t\right) = 
\pm 
\mu^{(\pm)}\left(\frac{\tau}{2}+t\right), 
\label{PSC10}
\end{align}
respectively. 
For the examples presented in this paper, 
we choose pulses where 
both the pulse-envelope function $\Omega(t)$ 
and the pulse-detuning function $\mu(t)$ are of 
positive parity. This entails that 
$\psi(t)$, according to (\ref{INTRO22}), 
has odd parity with respect to $\tau/2$, so that 
$\sin[\psi(t)]$  
is also of odd parity, resulting in a pulse function  
$g^{(-)}(t)$ of odd parity. 
Thus, 
to illustrate our pulse-generation method, we 
will in the following focus on negative-parity 
pulses, $g^{(-)}(t)$, constructed from a positive-parity 
pulse-envelope function $\Omega^{(+)}(t)$, 
negative-parity 
$\sin[\psi^{(-)}(t)]$, and positive-parity 
pulse-detuning function, $\mu^{(+)}(t)$. 
Since the  pulse function is of negative parity, 
we expand the pulse into a Fourier-sine 
series according to 
\begin{equation}
\label{eq:PSC11}
g^{(-)}(t) = \sum_{n=1}^{N_A} A_n 
\sin(2\pi n t/\tau), 
\end{equation}
where $A_n$, $n=1,\ldots,N_A$, are real expansion 
amplitudes and 
$N_A$ is chosen large enough to achieve 
convergence.  
The expansion \eq{PSC11} provides the additional 
benefit of
switching $g^{(-)}(t)$ off continuously at 
$t=\tau$ without a discontinuous jump to $g=0$ at 
$t=\tau$. It is straightforward to show that the 
expansion \eq{PSC11} is indeed odd with respect to 
$\tau/2$. 
The expansion \eq{PSC11} is complete, i.e., any pulse function 
$g^{(-)}(t)$ with $g^{(-)}(t=0)=g^{(-)}(t=\tau)=0$ can be 
represented this way.
Expanding the entire pulse 
$g^{(-)}(t)$ as a whole, and not $\Omega(t)$ and $\mu(t)$ 
separately, is natural, since neither 
$\Omega(t)$ nor $\mu(t)$ are known. 
In fact, expansion of the entire pulse function 
$g(t)$ is the key idea 
that motivated our method 
of AMFM pulse construction.

\subsection{Pulse construction}
\label{PG}

We focus in this section on computing 
the power-optimized pulse function $g^{(-)}(t)$ for a given 
set of motional-mode frequencies $\omega_p$ 
and Lamb-Dicke parameters 
$\eta_p^i$, $i=1,\ldots,N$, $p=1,\ldots,P$. 
Since in this case, according to 
(\ref{PSC8}), the $P$ equations 
$\int_0^{\tau}g^{(-)}(t)h_p^{(+)}(t)dt=0$ 
are automatically fulfilled, we need to 
fulfill, according to (\ref{PSC9}), 
only the set of equations 
\begin{equation}
\int_0^{\tau} g^{(-)}(t)h_p^{(-)}(t)\, dt = 0, 
\ \ \ p=1,\ldots, P. 
\label{PG1}
\end{equation}
Using the expansion \eq{PSC11} and the 
explicit form (\ref{PSC5}) of 
$h_p^{(-)}(t)$, we obtain the following set 
of real, linear equations 
\begin{equation}
\sum_{n=1}^{N_A} M_{pn} A_n = 0,\ \ \ p=1,\ldots,P , 
\label{PG2}
\end{equation}
where 
\begin{align}
M_{pn} &= \int_0^{\tau}\sin\left(2\pi n \frac{t}{\tau}\right) 
\sin\left[ \omega_p \left(\frac{\tau}{2}-t\right)\right]\, dt, 
\nonumber\\
p&=1,\ldots,P,\ \ \ n=1,\ldots,N_A. 
\label{PG3}
\end{align}
In matrix notation we may write (\ref{PG2}) in the form 
\begin{equation}
M \vec A = 0,
\label{PG4}
\end{equation}
where $M$ is the $P\times N_A$ coefficient matrix of 
(\ref{PG2}) and $\vec A$ is the amplitude vector of length $N_A$. 
In order for (\ref{PG4}) to have non-trivial solutions, we require 
$N_A > P$. 
In general, then, $M$ in (\ref{PG4}) will have rank $P$, 
and there exist 
$N_0=N_A-P$ non-trivial solutions 
$\vec A^{(\alpha)}$ of (\ref{PG4}), 
$\alpha=1,\ldots,N_0$.
Since $N_A>P$, the matrix $M$ is a rectangular matrix. This 
suggests to 
multiply (\ref{PG4}) from the left with the transpose, 
$M^T$, of $M$, 
which turns (\ref{PG4}) into 
the eigenvalue problem, 
\begin{equation}
\Gamma \vec A = 0, 
\label{PG5} 
\end{equation}
where $\Gamma=M^T M$ is a symmetric matrix, and  
we are looking for 
the $N_0$ eigenvectors $\vec A^{(\alpha)}$ 
of $\Gamma$ with eigenvalues 0. The $N_0$ nontrivial vectors 
$\vec A^{(\alpha)}$ with eigenvalues 0 span the kernel of 
the matrix $\Gamma$, also known as the 
{\it null space} of $\Gamma$. Numerically 
diagonalizing $\Gamma$, its eigenvalues typically are of the 
order of $10^{-12}$ in the null space, and several orders 
of magnitude larger in 
the complementary space. Thus, 
the transition from the null space to the complementary space 
is sharp, with eigenvalues jumping many orders of magnitude 
at the transition point. Therefore, the null space can be 
identified clearly and unambiguously. Without restriction of 
generality we may also assume that the null-space vectors are 
normalized. 
Since all null-space vectors $\vec A^{(\alpha)}$ have the common 
eigenvalue 0, the null space is degenerate. Thus, 
any linear combination of the $N_0$ null-space vectors 
$\vec A^{(\alpha)}$ are also null-space vectors, and we may 
assume that the $\vec A^{(\alpha)}$ form an orthonormal basis 
of the null space according to 
\begin{equation}
\vec A^{(\alpha)}\,^T \vec A^{(\beta)} 
= \delta_{\alpha\beta}, 
\label{PG6}
\end{equation}
where $\delta_{\alpha\beta}$ is the Kronecker symbol. 
Our goal now is to 
linearly combine the orthonormal null-space vectors 
$\vec A^{(\alpha)}$ with real expansion amplitudes 
$\Lambda_{\alpha}$ 
to find 
the optimal null-space vector 
\begin{equation}
\hat{\vec A} = \sum_{\alpha=1}^{N_0} \Lambda_{\alpha} 
\vec A^{(\alpha)} 
\label{PG7}
\end{equation}
such that 
\begin{equation}
\hat g^{(-)}(t) = \sum_{n=1}^{N_A} \hat A_n 
\sin \left(2\pi n \frac{t}{\tau} \right)
\label{PG8}
\end{equation}
is optimal in the sense that it produces 
$|\chi_{ij}|=\pi/8$, according to (\ref{INTRO4a}), and 
has the smallest possible norm 
\begin{equation}
\gamma^2 = 
|| \hat g^{(-)}(t) ||^2 = 
\frac{2}{\tau} 
\int_0^{\tau} 
\left[ \hat g^{(-)}(t) \right]^2 \, dt 
= \min_{\Lambda_{\alpha}} \sum_{n=1}^{N_A} 
\hat A_n^2, 
\label{PG9}
\end{equation}
which entails the smallest possible average 
power needed to execute a maximally entangling $\xxgate$ gate. 
Using (\ref{PG8}) with (\ref{METH3}) and (\ref{INTRO4a}) in 
\eq{INTRO4}, we obtain 
\begin{align} 
\frac{\pi}{8} &= \Bigg|
\sum_{p=1}^P \eta_p^i \eta_p^j 
\int_0^{\tau}\, dt_2\, \int_0^{t_2}\, dt_1 \, 
\nonumber\\
& \qquad
\hat g^{(-)}(t_2)\, \hat g^{(-)}(t_1)\, 
\sin\left[\omega_p(t_2-t_1)\right] \Bigg|
\nonumber\\
&= \left| \hat{\vec A\,}^T D \hat{\vec A} \right| , 
\label{PG10}
\end{align}
where $D$ is a real $N_A\times N_A$ matrix with matrix elements 
\begin{align}
D_{nm} &= \sum_{p=1}^P \eta_p^i \eta_p^j 
\int_0^{\tau}\, dt_2\, \int_0^{t_2}\, dt_1 \, 
\nonumber\\
&\sin \left(2\pi n \frac{t_2}{\tau} \right)
\sin\left[\omega_p(t_2-t_1)\right] 
\sin \left(2\pi m \frac{t_1}{\tau} \right) . 
\label{PG11}
\end{align} 
Since $\hat{\vec A}^{\;T} D \hat{\vec A}$ is a scalar, 
we can also write 
\begin{equation}
\hat{\vec A}^{\;T} D \hat{\vec A} = \frac{1}{2} \left[
\hat{\vec A}^{\;T} D \hat{\vec A} + 
\left(\hat{\vec A}^{\;T} D \hat{\vec A}\right)^T \right] 
= \hat{\vec A}^{\;T} S \hat{\vec A} , 
\label{PG12}
\end{equation} 
where 
\begin{equation}
S = \frac{1}{2} \left[ D + D^T \right] 
\label{PG13}
\end{equation} 
is a symmetric matrix. 
Using (\ref{PG13}) and (\ref{PG12}) in 
(\ref{PG10}) we now obtain 
\begin{equation}
\frac{\pi}{8} = \left|
{\vec \Lambda}^T R \vec \Lambda 
\right| , 
\label{PG13a}
\end{equation} 
where $\vec\Lambda$ is the vector of expansion 
amplitudes $\Lambda_{\alpha}$, 
$\alpha=1,\ldots,N_0$, and 
$R$ is the symmetric, reduced $N_0\times N_0$ matrix 
with matrix elements 
\begin{equation}
R_{\alpha\beta} = {\vec A}^{(\alpha)}\,^T S 
{\vec A}^{(\beta)},\ \ \ 
\alpha, \beta=1,\ldots,N_0. 
\label{PG14}
\end{equation} 
Since $R$ is symmetric, it can be diagonalized, 
\begin{equation}
R\, \vec V^{(k)} = \lambda_k\, \vec V^{(k)} , \ \ \ 
k=1,\ldots,N_0 , 
\label{PG14a}
\end{equation} 
where, since $R$ is real and symmetric, the 
eigenvectors $\vec V^{(k)}$ can be assumed orthonormal. 
We now linearly combine the vector of 
expansion amplitudes $\vec \Lambda$ from the 
set of vectors $\vec V^{(k)}$ according to 
\begin{equation}
\vec\Lambda = \sum_{k=1}^{N_0} v_k \vec V^{(k)}. 
\label{PG15}
\end{equation} 
According to (\ref{PG9}), we now have to determine the expansion 
amplitudes $v_k$ such that 
\begin{equation}
\gamma^2 = \min_{v_k} \hat{\vec A}^{\;T} \hat{\vec A} = 
\min_{v_k} \hat{\vec \Lambda}^T \hat{\vec \Lambda} = 
\min_{v_k} \sum_{k=1}^{N_0} v_k^2 
\label{PG16}
\end{equation} 
under the condition 
\begin{equation}
\frac{\pi}{8} = | {\vec \Lambda}^T R {\vec \Lambda} | = 
|\sum_{k=1}^{N_0} v_k^2 \lambda_k|. 
\label{PG18}
\end{equation} 
Geometrically, (\ref{PG14a}) is a principal-axis 
transformation, $\vec V^{(k)}$ are the $N_0$ principal 
directions of $R$ in 
the null space, (\ref{PG16}) is 
a $N_0$-dimensional sphere of radius 
$\gamma$, and (\ref{PG18}) is 
a $N_0$-dimensional conic section with 
principal axes $|\lambda_k|^{-1/2}$. Thus, 
geometrically speaking, we are looking for the 
smallest sphere that touches the conic section. 
This is obviously achieved if the sphere is 
inscribed in the conic section and just touches 
the conic section along the principal axis 
with the smallest length, i.e., the largest 
$|\lambda_k|$ 
Thus, our optimization problem 
is solved: The optimal pulse (\ref{PG8}) is constructed 
with the help of the amplitudes 
\begin{equation}
\hat {\vec A} = \sum_{\alpha=1}^{N_0} \Lambda_{\alpha}^{(k_{\rm max})}  
\vec A^{(\alpha)}, 
\label{PG19}
\end{equation} 
where $k_{\rm max}$ is the index of the 
eigenvalue $\lambda_k$ of (\ref{PG14a}) with the largest modulus 
$|\lambda_k|$, and 
\begin{equation}
\vec \Lambda^{(k_{\rm max})} = v_{k_{\rm max}} \vec V^{(k_{\rm max})} , 
\label{PG20}
\end{equation} 
where 
\begin{equation}
v_{k_{\rm max}} = 
\left( \frac{\pi}{8 |\lambda_{k_{\rm max}}|} 
\right)^{1/2}. 
\label{PG21}
\end{equation} 
To illustrate the method discussed in this section, 
we show in the main text in Figs.~4{\bf a} and {\bf b}
the optimal pulse 
$\hat g(t)$ obtained for $N=5$ ions and $P=5$ 
motional modes for mode frequencies and Lamb-Dicke 
parameters as shown in Tables~\ref{Tab1} and \ref{Tab2}, 
respectively. The pulse has a symmetric envelope function
and is amplitude as well as frequency modulated. 

\begin{table}
\caption{\label{Tab1} Motional-mode frequencies. }
\centering
\begin{tabular}{cccccc}
\hline\hline
 &$\omega_p/2\pi$ [MHz] \\
\hline
$p=1$  &  2.26870             \\
$p=2$  & 2.33944     \\
$p=3$  & 2.39955    \\
$p=4$  & 2.44820    \\
$p=5$  & 2.48038    \\
\hline\hline
\end{tabular}
\end{table}
%
%
\begin{table}
\caption{\label{Tab2} Lamb-Dicke parameters}
\centering
\begin{tabular}{cccccc}
\hline\hline
               &$p=1$         &$p=2$           &$p=3$        &$p=4$     &$p=5$ \\
\hline
$i=1$      & 0.01248     & 0.03474       & 0.06091     & 0.07149   & -0.04996  \\
$i=2$      & -0.05479    & -0.07263     & -0.03150   & 0.03406   & -0.05016  \\
$i=3$      & 0.08428     & -0.00002     & -0.05848     & -0.00021   & -0.05013  \\
$i=4$      & -0.05440    & 0.07306      & -0.03098     & -0.03459    & -0.04991  \\
$i=5$      & 0.01243     & -0.03514     & 0.06094       & -0.07163  & -0.04946  \\
\hline\hline
\end{tabular}
\end{table}
%
 
\subsection{Analytical lower bound of required peak pulse power}
\label{APPCS} 
 In this section we derive an exact, closed-form, 
 integral-free, analytical expression for the lower bound 
 of the minimally required pulse power 
 needed to operate an $i\leftrightarrow j$ 
 XX gate. To be specific, throughout this section we 
 choose $P=N$, which is the mode in which 
 our quantum computer is operated \cite{IonQ}. 
 Generalizing the lower bound to the case 
 $P\neq N$ is straightforward. 
  
 We define 
 \begin{equation}
 G = \int_0^{\tau} g^2(t)\, dt  = \int_0^{\tau} \Omega^2(t)\sin^2[\psi(t)]\, dt 
 \leq \Omega_{\rm max}^2 \sigma ,
 \label{APPCS1}  
 \end{equation}
 where we defined 
 \begin{equation}
 \sigma = \int_0^{\tau} \sin^2[\psi(t)]\, dt. 
  \label{APPCS1a}  
 \end{equation}
 We also define 
 \begin{align}
 D 
 &=\sum_{pp'=1}^N 
 \eta_p^{i}  \eta_p^{j}  \eta_{p'}^{i}  \eta_{p'}^{j} 
 \int_0^{\tau} dt_2 \int_0^{t_2} dt_1\,
 \nonumber \\
 &\quad
 \sin[\omega_p(t_2-t_1)] 
 \sin[\omega_{p'}(t_2-t_1)] 
 \nonumber \\
 &=\frac{1}{4} \sum_{p=1}^N (\eta_p^{i}\eta_p^{j})^2 
 \left[ \tau^2 - \frac{1}{\omega_p^2}\sin^2(\omega_p\tau) \right] +
 \nonumber \\
 &\quad 
 \sum_{p\neq p'=1}^N 
 \eta_p^{i}  \eta_p^{j}  \eta_{p'}^{i}  \eta_{p'}^{j} 
 \left\{ 
 \frac{1}{(\omega_p-\omega_{p'})^2} \sin^2\left[\left( 
 \frac{\omega_p-\omega_{p'}}{2}\right)\tau \right] \right.
 \nonumber \\
 &\quad\quad
 \left.
 - 
 \frac{1}{(\omega_p+\omega_{p'})^2} \sin^2\left[\left( 
 \frac{\omega_p+\omega_{p'}}{2}\right)\tau \right] 
\right\} 
\leq \frac{\tau^2}{4} \beta^4, 
 \label{APPCS2}  
 \end{align}
 where we defined 
 \begin{equation}
 \beta = \left[ \sum_{p=1}^N (\eta_p^{i}\eta_p^{j})^2  
 + \sum_{p\neq p'=1}^N  
 \frac{4
 |\eta_p^{i}  \eta_p^{j}  \eta_{p'}^{i}  \eta_{p'}^{j} |} 
 {(\omega_p\tau - \omega_{p'}\tau)^2 } \right]^{1/4} , 
 \label{APPCS2beta}  
 \end{equation}
which, for fixed $N$, is essentially 
a constant, which depends only weakly on $\tau$, 
i.e., 
\begin{equation}
\beta(\tau)\sim \left[
\sum_{p=1}^N (\eta_p^{i}\eta_p^{j})^2 \right]^{1/4} . 
\label{beta1}
\end{equation}
For instance, 
for an $80\,\mu$s pulse, and 
the mode frequencies and $\eta$ values listed 
in Tables~\ref{Tab1} and \ref{Tab2}, 
respectively, the first term in (\ref{APPCS2beta}) 
is $2\times 10^{-5}$ while the second term is 
$5\times 10^{-8}$. Therefore, in practice, 
the second term in (\ref{APPCS2beta}) may be 
neglected. Since the Lamb-Dicke parameters 
$\eta_p^j$ are proportional to the $j$th component 
of a unit vector \cite{LD}, 
we have, on average, 
$\eta_p^j\sim 1/\sqrt{N}$, which then, because 
of (\ref{beta1}), implies 
\begin{equation}
\beta \sim 1/N^{1/4}. 
\label{beta2}
\end{equation}
With these definitions, and using the 
Cauchy-Schwarz 
inequality for integrals, we obtain: 
 \begin{align} 
 \frac{\pi}{8} &= 
 \chi_{i,j} \nonumber \\
 &= 
 \left|
 \sum_{p=1}^N\, \eta_p^{i}  \eta_p^{j} 
 \int_0^{\tau} dt_2 \int_0^{t_2} dt_1\, 
 g(t_2) g(t_1) 
 \sin[\omega_p(t_2-t_1)] \right| 
 \nonumber \\
 &\leq \left[   \int_0^{\tau} dt_2 \int_0^{t_2} dt_1\,  
 g^2(t_2) g^2(t_1) \right]^{1/2} 
 \nonumber\\
 &\left\{ 
  \int_0^{\tau} dt_2 \int_0^{t_2} dt_1\,  
  \left(    \sum_{p=1}^N\, \eta_p^{i}  \eta_p^{j}  
     \sin[\omega_p(t_2-t_1)]   \right)^2\ \right\}^{1/2}   
 \nonumber \\
&= \left[ \frac{1}{2} 
 \int_0^{\tau} dt_2 \int_0^{\tau} dt_1\, 
  g^2(t_2) g^2(t_1) \right]^{1/2} \ D^{1/2} 
 \nonumber \\
  &= \frac{1}{\sqrt{2}} G D^{1/2} \leq 
  \frac{\tau\sigma}{2\sqrt{2}} \Omega_{\rm max}^2 \beta^2 . 
\label{APPCS3}  
\end{align}
Using $\sin^2[\psi(t)] \leq 1$, 
which is valid 
for all arguments $\psi(t)$, 
the most straightforward, exact 
estimate for $\sigma$ is 
\begin{equation}
\sigma \leq \tau .
\label{APPCS3e}  
\end{equation}
Using this in the inequality (\ref{APPCS3}) 
and solving for $\Omega_{\rm max}$, we obtain 
\begin{equation}
\Omega_{\rm max} \geq \frac{\sqrt{\pi}}{2^{3/4}\tau\beta} 
\label{APPCS5a}  
\end{equation}
or, transitioning to lab frequency, 
\begin{equation}
f_{\rm max} \geq \frac{1}{2^{7/4}\sqrt{\pi} \tau \beta} . 
\label{APPCS5b}  
\end{equation}
This is the formula used to compute 
the analytical lower bounds 
of minimally required 
power to operate an XX gate, 
stated in the lower half of Table~\ref{Tab3}. 
The lower bound (\ref{APPCS5a}) 
[(\ref{APPCS5b}), respectively] 
 is an important result. 
Since all the steps leading to (\ref{APPCS5a}) 
[(\ref{APPCS5b}), respectively] 
are 
rigorous, the lower bound 
(\ref{APPCS5a}) 
[(\ref{APPCS5b}), respectively] 
implies that no 
pulse exists, even in principle, that would require 
lower power than indicated by (\ref{APPCS5a}) 
[(\ref{APPCS5b}), respectively] 
to operate an XX gate. 
We also see that, because of (\ref{beta2}), 
$\Omega_{\rm max}$ ($f_{\rm max}$, respectively) 
scales like $\sim N^{1/4}$. 
 
In many cases (\ref{APPCS5a}) 
[(\ref{APPCS5b}), respectively] 
may be sharpened if 
lower ($\mu_{\rm min}$) and upper ($\mu_{\rm max}$)
bounds for the detuning function 
$\mu(t)$ are available (see, e.g., 
the main text Fig.~\ref{fig:AF}{\bf b}
), i.e., 
 \begin{equation}
 \mu_{\rm min} \leq \mu(t) \leq \mu_{\rm max}, \ \ \ t\in[0,\tau]. 
 \label{APPCS2a}  
 \end{equation} 
 We define 
 \begin{equation}
 \psi_{\tau} = \int_0^{\tau} \mu(t)\, dt \leq \mu_{\rm max} \tau . 
  \label{APPCS2b}  
 \end{equation} 
Since $\psi(t)$, according to (\ref{INTRO2}), 
is defined via an integral, 
and since $\mu(t)>0$ for all $t$, 
$\psi(t)$ is 
a monotonically increasing function of $t$. 
Therefore, in (\ref{APPCS1a}), 
we may change
variables from $t$ to $\psi$ to obtain 
\small
\begin{align}
\sigma &= \int_{\psi_0}^{\psi_0+\psi_{\tau}} 
\sin^2(\psi) \frac{1}{\mu[t(\psi)]}\, d\psi 
\leq \frac{1}{\mu_{\rm min}} \int_{\psi_0}^{\psi_0+\psi_{\tau}}  
\sin^2(\psi) \, d\psi 
\nonumber \\ 
&= \frac{1}{2\mu_{\rm min}}\left[ 
\psi_{\tau} - \cos(2\psi_0+\psi_{\tau})\sin(\psi_{\tau}) \right]
\leq \frac{1}{2\mu_{\rm min}} (\psi_{\tau}+1) 
\nonumber \\ 
&\leq  \frac{1}{2\mu_{\rm min}} (\mu_{\rm max}\tau+1) , 
\label{APPCS3b}  
\end{align}
\normalsize
where, in the last inequality, we used (\ref{APPCS2b}). 
With (\ref{APPCS3b}), the inequality (\ref{APPCS3}) can now 
be stated in the form 
\begin{equation}
\frac{\pi}{8} \leq \frac{\Omega_{\rm max}^2}
{4\sqrt{2}\mu_{\rm min}} (\mu_{\rm max}\tau+1) \tau\beta^2, 
\label{APPCS4}  
\end{equation}
or, solved for $\Omega_{\rm max}$, 
\begin{equation}
\Omega_{\rm max} \geq \frac{1}{2^{1/4}\tau\beta} 
\sqrt{\frac{\pi\mu_{\rm min}}{\mu_{\rm max}+1/\tau}}. 
\label{APPCS5}  
\end{equation}
Transitioning from angular frequency to lab frequency in Hz, 
we obtain 
\begin{equation}
f_{\rm max} = \frac{\Omega_{\rm max}}{2\pi} \geq 
 \frac{1}{2^{5/4}\sqrt{\pi}\tau\beta} \sqrt{\frac{\mu_{\rm min}}
{\mu_{\rm max} + 1/\tau}} .
\label{APPCS6}  
\end{equation}
This is our central result. 
No pulse exists with a power 
lower than stated in (\ref{APPCS6}) if $\chi_{ij}$ 
is determined by 
(\ref{eq:chi}). 
 
To illustrate our analytical result, we show in Table~\ref{Tab3} 
a comparison between our analytical lower limit of peak pulse 
power and numerically obtained peak pulse powers for 
our sample case of $N=5$ ions and $P=5$ 
motional-mode frequencies as 
listed in Tables~\ref{Tab1} and \ref{Tab2}. 
We see that our analytical result is indeed lower than 
all numerically obtained peak pulse powers, but that both 
are qualitatively close. 
%
%
\begin{table}
\caption{\label{Tab3} Analytical lower bounds of 
minimally required analytically computed peak power (lower triangle) and 
numerically computed peak power of optimal pulses 
(upper triangle) for gate combinations 
$i \leftrightarrow j$, mode frequencies 
as listed in Table~\ref{Tab1}, Lamb-Dicke parameters 
$\eta$ as listed in Table~\ref{Tab2}, and $\tau=300\,\mu$s. Powers quoted are in kHz. 
Basis size: $N_A=1000$; $n_{\rm min}=1$. }
\centering
\begin{tabular}{cccccc}
\hline\hline
 &$j=1$ &$j=2$ &$j=3$ &$j=4$  &$j=5$ \\
\hline
$i=1$  & *             & 37.8     & 28.9   & 43.6   & 25.7  \\
$i=2$  & 8.09   & *             & 25.6   & 23.5   & 43.7  \\
$i=3$  & 8.35   & 7.49   & *             & 25.7   & 28.9  \\
$i=4$  & 8.09   & 6.80   & 7.52    & *            & 37.0  \\
$i=5$  & 6.73   & 8.09   & 8.36    & 8.08  & *  \\
\hline
\end{tabular}
\end{table}
%
\subsection{Power and execution time scaling}
\label{SCAL} 
The execution time of our 
linear pulse-construction algorithm is dominated by two diagonalizations, 
i.e., the diagonalization of the matrix $\Gamma=M^T M$ 
[see (\ref{PG5})]  
and the reduced matrix $R$ 
[see (\ref{PG14})]. 
The dimension of $\Gamma$ is $N_A\times N_A$, and 
the dimension of $R$ is $(N_A-P)\times (N_A-P)$. 
Therefore, the execution time of our algorithm scales like 
$\sim N_A^3$.  
Since, in general, 
$N_A\gg P$, the execution time 
is dominated by $N_A$ and depends on $P$ only via 
$N_A>P$, which is needed 
for a nontrivial null space. Therefore, 
the overall scaling is dominated by $N_A$ and 
the algorithm scales 
like $\sim N_A^3$. 
We confirmed the $\sim N_A^3$ scaling of 
our algorithm in numerous pulse-generation runs. 
 
We also investigated the scaling of pulse power in $N$ 
with up to $N=50$ ions. For our 
investigation of power scaling we generated motional-mode 
frequencies and 
Lamb-Dicke parameters according to 
the procedure outlined in 
\cite{MMETA}. 
We used simulated ion positions, approximately equi-spaced 
with a spacing of about $5\,\mu$m and a frequency ratio 
of axial to radial trap frequencies of 
$\omega_x/\omega_r=0.088$. We focused on operating 
an XX gate between ions 1 and 3. 
For these 
parameters and for $N=50$ particles we obtained 
an average motional-mode frequency 
spacing of $\Delta f=1.46\,$kHz. 
We found that our algorithm is stable only if 
$\tau\Delta f\approx 1$. Therefore, for our 
power-scaling simulations, we chose 
$\tau=500\,\mu$s. 
The result of our power-scaling 
simulations in a basis of $N_A=1000$ states is shown 
in the main text Fig.~2{\bf c}. 
We see that the power scales approximately 
like 
$N^{1/4}$, which is consistent with 
the analytical 
power scaling 
(\ref{APPCS5a}) with 
(\ref{beta2}) 
(gray full line in Fig.~2{\bf c}). 
As pointed out in the main text, 
knowledge of power scaling is important 
since, apart from possibly damaging 
optical components when applying too 
much power, increasing power also 
enhances important sources of errors. 
 
\subsection{Power optimality requires identical pulses}
\label{POSP} 
In this section we show that if we do not actively stabilize 
the degree of entanglement $\chi$, a gate is power 
optimal if ions $i$ and $j$ participating in a two-qubit gate 
are illuminated with identical laser pulses, i.e., 
$g_i(t)=g_j(t)=g(t)$. To show this, let 
$\vec A$ and $\vec B$ be the expansion amplitudes 
of $g_i$ and $g_j$, respectively. 
Then, the degree of entanglement is 
$\chi = \vec A^T R \vec B$, 
where the symmetric matrix $R$ is defined in 
(\ref{PG14}). Define 
$P_A^2 = \vec A^T\vec A$ and 
$P_b^2 = \vec B^T\vec B$. Then, the task 
is to minimize $P^2=P_A^2+P_B^2$ under the constraint 
$\chi$. Thus, the target function to be minimized is 
$F(\vec A,\vec B)=P^2-\lambda \chi$, where 
$\lambda$ is a Lagrangian parameter. This 
yields two equations: 
\begin{align}
    \frac{\partial F}{\partial \vec A} = 2\vec A-
    \lambda R \vec B = \vec 0\ \ \ \Rightarrow\ \ \ 
    \vec A = \frac{1}{2} \lambda R \vec B, 
\label{SYM1}\\
    \frac{\partial F}{\partial \vec B} = 2\vec B-
    \lambda R \vec A = \vec 0,\ \ \ \Rightarrow\ \ \
        \vec B = \frac{1}{2} \lambda R \vec A. 
\label{SYM2}
\end{align}
From (\ref{SYM1}) and (\ref{SYM2}) we obtain 
immediately 
$P_A^2=\lambda \vec A^T R\vec B/2=\lambda\chi/2$ and 
$P_B^2=\lambda \vec B^T R\vec A/2=
\lambda \vec A^T R \vec B / 2 = \lambda\chi/2$. 
Thus, $P_A=P_B$, i.e., for power optimality the 
same power must be directed at both ions. 
From 
From (\ref{SYM1}) and (\ref{SYM2}) we further obtain 
$\vec A=(\lambda R/2)(\lambda R\vec A/2)=\lambda^2R^2\vec A/4$ 
and 
$\vec B=(\lambda R/2)(\lambda R\vec B/2)=\lambda^2R^2\vec B/4$, 
i.e., 
$\vec A$ and $\vec B$ satisfy the same eigenvalue equation. 
This means that, up to normalization, $\vec A$ and 
$\vec B$ are the same. Together with $P_A=P_B$, we 
now have $g_i(t)=g_j(t)=g(t)$. 


\subsection{Stabilization against mode-frequency fluctuations}
\label{Stab}

In this section we show that our linear approach lends itself 
naturally to a method of constructing pulses that stabilize 
the fidelity of the $\xxgate$ gate against mode drifts and mode fluctuations. 
Due to uncontrollable effects, such as stray electromagnetic fields, 
build-up of charge in the trap due to photoionization  
or temperature fluctuations, 
the frequencies of the motional modes, 
$\omega_p$, will drift or 
fluctuate in time. Therefore, in a typical quantum-computer run, 
one would determine the current values of $\omega_p$ and the 
associated pulse $\hat g(t)$. However, 
typically over a timespan of minutes, the 
motional-mode frequencies $\omega_p$ will drift 
with typical excursions 
of $\Delta\omega_p/(2\pi) \approx 1\,$kHz. 
If we now use $\hat g(t)$, determined on the 
basis of the original mode 
frequencies $\omega_p$, in the situation of the 
drifted modes, $\omega_p+\Delta\omega_p$,
the set of equations \eq{INTRO1} are 
no longer fulfilled, 
resulting in a reduction of 
the fidelity of the $\xxgate$ gate. 
A simple estimate for the infidelity increase 
due to the now 
non-zero $\alpha$'s in \eq{INTRO1} is presented in 
\cite{GREENPAP}. According to \cite{GREENPAP}, 
at zero temperature of the 
motional-mode phonons, 
the infidelity, $\hat F$, 
is approximately given by 
\begin{equation}
\hat F = \frac{4}{5}\sum_p 
\left(|\alpha_{i,p}|^2 + |\alpha_{j,p}|^2 \right). 
\label{MOM1}
\end{equation}
This suggests stabilizing the fidelity of the 
quantum computer against mode drifts and fluctuations 
by requiring that $\alpha_{ip}$ be stationary up to 
$n$th order with respect to variations in $\omega_p$. 
This is easily accomplished by adding the following 
set of equations to the set of equations \eq{INTRO1}: 
\begin{align}
&\frac{\partial^k \alpha_{ip}}{\partial \omega_p^k} = 0 = 
\int_0^{\tau} (it)^k \Omega(t) \sin[\psi(t)]
e^{i\omega_p t}\, dt, 
\nonumber\\
&i=1,\ldots,N,\ \ \ p=1,\ldots,P,\ \ \ k=1,\ldots,K. 
\label{MOM2}
\end{align}
Because of the presence of the factor 
$t^k$ in the integrand of (\ref{MOM2}), we call 
this extension of our linear approach the 
{\it moments approach}. 
Adding the moments equations (\ref{MOM2}) to 
the set \eq{INTRO1} 
does not change the 
linearity of our method. 
The same techniques can be 
applied in solving this extended system of 
linear equations as was described in 
the main text. 

\subsection{Demodulation of pulses}
\label{denom}

The optimal pulse functions $\hat g(t)$ are simultaneously amplitude-, frequency-, and phase-modulated pulses.  
In this section we show how to demodulate the pulse 
$\hat g(t)$, i.e., how to separate $\hat g(t)$ into its amplitude function 
$\Omega(t)$ and its detuning function $\mu(t)$.  

The first step of our demodulation procedure is to find 
the zeros $\zeta_j$ of $\hat g(t)$. This is numerically unproblematic, 
since the detuning function $\mu(t)$ is bounded away from zero, 
which means that degeneracies of nontrivial zeros ($\zeta_j>0$) do not occur. 
In addition, in numerous simulation runs, we observed that 
the envelope function $\Omega(t)$ was always bounded away 
from zero. Therefore, 
in order not to complicate the discussion, we may also assume 
that $\Omega(t)$ does not have any zeros. Thus, all the zeros 
in $\hat g(t)$ are caused by zeros of $\sin[\psi(t)]$, i.e., 
$\psi(\zeta_j)$ is a multiple of $\pi$. Since no degenerate zeros 
occur, we have even more, namely 
\begin{equation}
\psi(\zeta_j) = j\pi,\ \ \ j=0,1,\ldots,N_z-1, 
\label{DEM1}
\end{equation}
where $N_z$ is the total number of zeros of $\hat g(t)$, 
including the zero $\zeta_0=0$ at $t=0$ and 
$\zeta_{N_z-1}=\tau$ at $t=\tau$. 
We now approximate the detuning function $\mu(t)$ as 
a constant between zeros of $\hat g(t)$, i.e., 
\begin{equation}
\mu(t)\approx \mu_j, \ \ \ \zeta_{j-1}<t<\zeta_j,\ \ \ 
j=1,2,\ldots,N_z - 1. 
\label{DEM2}
\end{equation}
With (\ref{INTRO2}) and (\ref{DEM1}) this entails 
\begin{align}
&\psi(\zeta_{j})-\psi(\zeta_{j-1}) = 
\int_{\zeta_{j-1}}^{\zeta_j} \mu(t')\, dt' = 
\mu_j (\zeta_j-\zeta_{j-1}) = \pi 
\nonumber \\ 
&\implies \ \ \ \mu_j = \frac{\pi}{\zeta_j-\zeta_{j-1}}, \ \ \ 
j=1,2,\ldots,N_z - 1. 
\label{DEM3}
\end{align}
As an example of frequency demodulation, 
the main text Fig.~\ref{fig:AF}{\bf b}
shows the result of the detuning 
function $\mu(t)$ for the pulse shown in 
the main text Fig.~\ref{fig:AF}{\bf a}.
We see that $\mu(t)$ hovers about the middle motional mode, 
staying away from 
the strongly 
heating mode with the 
highest motional frequency. 
Since $g(t)$, for the example 
shown in Fig.~\ref{fig:AF}{\bf a}, has a dense set of $N_z=387$ zeros,  
$\mu(t)$ approximated by as many piece-wise constant plateaus appears as a smooth 
function on the scale of Fig.~\ref{fig:AF}b.
 
We now turn to extracting the pulse envelope function 
$\Omega(t)$ from $\hat g(t)$. Differentiating 
(\ref{g-decomposed})
and evaluating the result at the zeros $\zeta_j$ of $\hat g(t)$ 
yields 
\begin{align}
\hat g'(\zeta_j) &= \Omega'(\zeta_j)\sin[\psi(\zeta_j)] 
+ \Omega(\zeta_j)\cos[\psi(\zeta_j)] \psi'(\zeta_j) 
\nonumber \\ 
&= (-1)^j \Omega(\zeta_j)\mu(\zeta_j), 
\label{DEM3a}
\end{align}
where we used (\ref{INTRO22}) and (\ref{DEM1}). This 
equation can be solved for $\Omega(\zeta_j)$ with the result 
\begin{equation}
\Omega(\zeta_j) =  (-1)^j \sigma \frac{\hat g'(\zeta_j)}{\mu(\zeta_j)}, 
\ \ \ j=1,\ldots,N_z-1,  
\label{DEM4}
\end{equation}
where we inserted the factor 
$\sigma=-\hat g'(\zeta_1)/|\hat g'(\zeta_1)|$, which ensures 
that $\Omega(t)$ is ``right-side up'', i.e., 
if it does not change sign, $\Omega(t)>0$ for all $t$. 
Since $\hat g(t)$, according to (\ref{PG8}), is represented 
by a Fourier series, it is trivial to obtain 
\begin{equation}
\hat g'^{(-)}(t) = \frac{2\pi}{\tau} \sum_{n=1}^{N_A} n \hat A_n 
\cos \left(2\pi n \frac{t}{\tau} \right)
\label{DEM5}
\end{equation}
and thus $\hat g'^{(-)}(\zeta_j)$. The values of the 
detuning function $\mu(\zeta_j)$ may be obtained in 
several ways. We may use spline interpolation of 
the data set of values $\mu_j$ as defined in 
(\ref{DEM3}), or, as we found, with sufficient accuracy, 
simply use 
(i) $\mu(\zeta_j)=\mu_j$, 
(ii) $\mu(\zeta_j)=\mu_{j+1}$, or 
(iii) $\mu(\zeta_j)=(\mu_{j+1}+\mu_j)/2$. 
We used method (i) to obtain the pulse envelope 
function $\Omega(t)/(2\pi)$ 
(heavy orange line in the main text Fig.~\ref{fig:AF}{\bf a})
of the pulse $\hat {g}^{(-)}(t)$, shown as the 
thin green line in the main text Fig.~\ref{fig:AF}{\bf a}.
The main text Fig.~\ref{fig:AF}{\bf a}
shows that our 
amplitude demodulation technique presented above works very 
well and accurately extracts the envelope function. 

At this point 
we may wonder how well the exact pulse 
$\hat {g}^{(-)}(t)$ is approximated by 
the pulse $\tilde {g}^{(-)}(t)$, i.e., the pulse 
reconstructed 
via (\ref{METH3}) from the amplitude 
and detuning functions obtained by 
demodulating $\hat {g}^{(-)}(t)$ according to 
the above procedures. Therefore, 
to get a first impression of the 
accuracy of our pulse demodulation method, we compute 
\begin{equation}
\Delta g^2 = \frac{1}{\tau} \int_0^{\tau}
\left[ \hat g^{(-)}(t) - \tilde g^{(-)}(t) \right]^2\, dt , 
\label{DEM6}
\end{equation}
where, for $j=1,2,\ldots,N_z-1$, 
\begin{equation}
\tilde g^{(-)}(t) = \Omega_j \sin[\psi_{j-1}+\mu_j(t-\zeta_{j-1})], 
\ \ \ \zeta_{j-1}\leq t < \zeta_j, 
\label{DEM7}
\end{equation}
\begin{equation}
\Omega_j = (-1)^j \frac{\hat g'(\zeta_j)}{\mu_j},
\label{DEM8}
\end{equation}
and 
\begin{equation}
\psi_j = \psi_{j-1}+\mu_j(\zeta_j-\zeta_{j-1}). 
\label{DEM9}
\end{equation}
Notice that $\Omega_j$ in (\ref{DEM8}) does not contain 
the factor $\sigma$ as in (\ref{DEM4}), since this time 
we do not need the ``right-side up'' pulse, but the pulse 
that has the same sign of the amplitude as 
$\hat {g}^{(-)}(t)$. 
For the example shown in 
the main text Fig.~\ref{fig:AF}{\bf a},
we obtain 
$\Delta g^2= 1.3\times 10^{-5}$. 
Hence, the 
pulse $\tilde {g}^{(-)}(t)$ 
reconstructed from the demodulated 
pulse $\hat {g}^{(-)}(t)$ 
is sufficiently accurate to guarantee 
high-fidelity gates.

\subsection{Stabilization against pulse-timing errors}
\label{tau-stab} 
Once $\tau$ and the mode frequencies $\omega_p$ 
are given, our method determines the amplitudes 
$A_n$, which are then fixed when implementing the 
pulse $g(t)$. However, the experimental clock 
may run fast or slow, which results in pulse-timing 
errors. To stabilize against pulse-timing 
errors of this nature, we require, 
for all $l=0,\ldots,L$ and $p=1,\ldots,N$: 
\begin{align}
    &\frac{\partial^l}{\partial \tau^l}
    \int_0^{\tau} g(t) e^{i\omega_p t}\, dt  
    \nonumber \\ 
&=\frac{\partial}{\partial \tau^l} \int_0^{\tau}
\sum_n A_n\sin(2\pi n t/\tau) e^{i\omega_p t}\, dt  
\nonumber \\ 
&=\sum_n A_n \left[ \frac{\partial}{\partial \tau^l}
\int_0^{\tau} \sin(2\pi n t/\tau) e^{i\omega_p t}\, dt 
\right] 
\nonumber \\ 
&=\sum_n A_n Q_{np}^{(l)} . 
\label{PTE1}
\end{align}
Since $L=0$, according to (\ref{eq:alpha}) of the main text, 
is already satisfied, (\ref{PTE1}) represents 
$LN$ linear equations that may be added to 
the $N(K+1)$ linear equation of motional-mode 
stabilization. Apart from providing explicit 
formulas for stabilization against pulse-timing 
errors, this also provides a template for  
stabilizing against other types of errors and 
parameter fluctuations, and shows that it is 
straightforward to extend our method by any 
number of such linear constraints, as long 
as the dimension of the available null-space 
is not exhausted.

\subsection{Implementation details}
\label{ImpDet}

In this section, we present the pulse-level implementation details. 
Figure~\ref{fig:impPulseShapes} shows the amplitude and frequency 
profiles of pulses, 
represented according to equation 
(\ref{g-decomposed}) in the main text, 
with moment stabilization orders $K=1,4,7$, 
used to implement the $\xxgate$ gates on our 5-qubit, 7-ion TIQIP. 
The motional-mode frequencies for each of the 
experiments are reported in 
Tables~\ref{tab:ModeSpectrum1} and \ref{tab:ModeSpectrum2}. The two-qubit gates were performed on qubits 3 and 4, with indexing starting at 0. 
 
We note that the 
even-parity population, used 
in Fig.~\ref{fig:exp} to demonstrate
stabilization against mode-frequency 
drift, 
is not a complete 
characterization of the 
resulting quantum gate 
in terms of fidelity. 
However, the even-parity population 
is the most relevant metric to use when evaluating the general 
experimental performance 
of our methods. 
The formalism for stabilizing 
phase space closure 
at the end of the 
gate with respect to 
gate frequency offset, 
relevant to Fig.~\ref{fig:exp}, 
aims to minimize 
$|\alpha_p|$ in 
(\ref{eq:alpha}), which is most closely 
related to the odd-parity population, 
i.e., $1$ minus the 
even-parity population. 
Moreover, fidelity  
contains within itself other 
sources of error accumulated 
during the parity contrast 
measurement (e.g., intensity 
noise differential, 
phase stability, laser-beam steering, 
overlap errors of the lasers with the 
positions of the trapped ions, etc.), which would mask 
the effect of stabilization 
against motional-mode 
drift. 
Thus, 
we decided to present the 
even-parity population as the 
most relevant quantity of 
interest to our work.

\begin{figure*}[]
    \centering
    \includegraphics[width=\linewidth]{FigS1.pdf}
    \caption{Amplitude and frequency profiles of pulses used in the implementation of the power optimal, stabilized entangling two-qubit gates. The gate time $\tau$ for each of the pulses is $\tau$ $\approx$ 550.0$\mu$s.}
    \label{fig:impPulseShapes}
\end{figure*}

\begin{table}[]
\begin{tabular}{c}
Mode Frequencies (MHz)\\ \hline \hline
    2.692 \\ 
    2.728 \\
    2.765 \\
    2.801 \\
    2.834 \\
    2.866 \\
    2.877 \\
    \hline
\end{tabular}
\caption{Mode frequencies of the motional modes of our 7-ion chain for the $K=1$ and $K=7$ gates shown in Fig.~\ref{fig:impPulseShapes}. }
\label{tab:ModeSpectrum1}
\end{table}
\begin{table}[]
\begin{tabular}{c}
Mode Frequencies (MHz)\\ \hline \hline
    2.690 \\ 
    2.726 \\
    2.763 \\
    2.799 \\
    2.832 \\
    2.864 \\
    2.876 \\
    \hline
\end{tabular}
\caption{Mode frequencies of the motional modes of our 7-ion chain for the $K=4$ gate shown in Fig.~\ref{fig:impPulseShapes}. }
\label{tab:ModeSpectrum2}
\end{table}


\subsection{Fixed-detuning step pulses} 
\label{STP} 
Possibly the most widely studied type of 
fixed-detuning pulses are segmented step pulses 
\cite{AM}. According to this method, 
the detuning function $\mu(t)$ is set to a constant, i.e., 
$\mu(t)=\mu_0={\rm const}$ for $0\leq t\leq \tau$, 
and the pulse interval $[0,\tau]$ is broken up into 
$N_{\rm seg}>P$ equi-spaced intervals 
$[t_{j-1},t_j]$, $t_0=0$, 
$t_j=j\Delta t$, 
$\Delta t = \tau/N_{\rm seg}$, 
$j=1,\ldots,N_{\rm seg}$, in which the 
pulse amplitude is set to a constant, i.e., 
\begin{equation}
\Omega(t) = \Omega_j,\ \ \ {\rm for}\ t_{j-1}\leq t\leq t_j,\ \ \ 
j=1,\ldots,N_{\rm seg}.  
\label{STP1} 
\end{equation}
For this type of pulses our methods 
are directly applicable with 
only two minor modifications. 
(i) For given $\mu_0$, we choose the gate length $\tau$ such 
that $J=\mu_0\tau/\pi$ is an integer. This way, 
still requiring that $\Omega(t)$ is an even function 
with respect to $\tau/2$, we obtain even- or odd-parity 
pulses, 
$\hat g^{(\pm)}(t) = \Omega(t) \sin(\mu_0 t)$, 
for $J$ odd or even, respectively. Since $J$ needs to be an 
integer to obtain the desired symmetry 
classes, $\tau$ can take only discrete values. 
However, since for quantum 
computer hardware of practical 
interest (for instance, Yb-ion quantum 
computers \cite{UMD}), 
the detuning $\mu_0$ is such that $J$ is a large
integer (of the order of 1000), the 
discretization of $\tau$ is of no consequence in practice. 
(ii) The second modification 
concerns the computation of the matrix $M$ 
defined in (\ref{eq:alpha}). 
For step pulses, we let $M_{pn}\rightarrow M_{pj}$, 
where, including both negative- and positive-parity pulses, 
we have 
\begin{equation}
M^{(\pm)}_{pj} = \int_{t_{j-1}}^{t_j}\, \sin(\mu_0 t) 
h_p^{(\pm)}(t)\, dt. 
\label{STP2} 
\end{equation}
Defining $\vec A = (\Omega_1,\Omega_2,\ldots,\Omega_{N_{\rm seg}})$, 
all the procedures outlined in Section \ref{PG}
can now be 
applied to construct step pulses.  
 
%
\begin{figure}
\includegraphics[scale=0.7,angle=0]{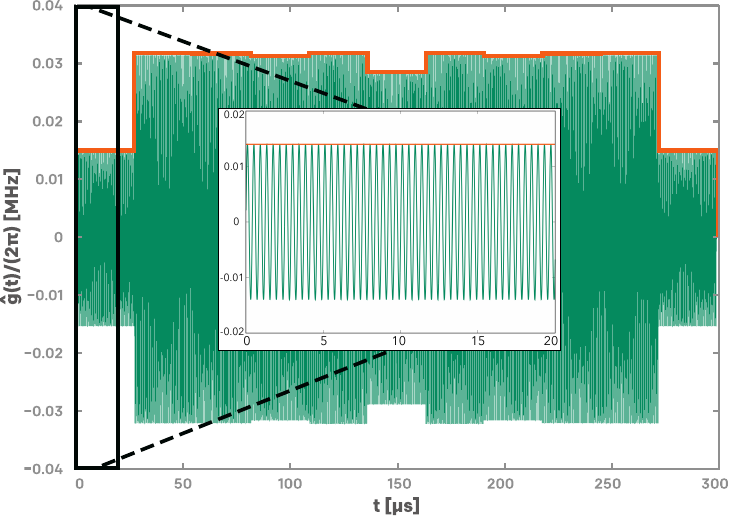}
\caption {
Fixed-detuning 
step pulse for $N=5$, 
$N_{\rm seg}=11$, $J=1434$, $\mu_0/(2\pi)=2.396\,$MHz, 
which corresponds to $\tau\approx 299.26\,\mu$s. 
The thin green line is the step pulse $\hat g^{(-)}(t)$;  
the thick, orange line is the piecewise constant 
pulse envelope function 
$\Omega(t)/(2\pi)$.  
      }
\label{fig:figstep}
\end{figure}
%
 
\Fig{figstep} shows an example of a 
negative-parity step pulse, 
generated for the same set of motional-mode frequencies 
and Lamb-Dicke parameters as in 
the main text Fig.~\ref{fig:AF}{\bf a}.
Although \fig{figstep} shows the negative-parity pulse with the 
lowest peak-power requirement that we found in the 
detuning interval from $\mu_0/(2\pi)=2.2\,$MHz to 
$\mu_0/(2\pi)=2.6\,$MHz, we see that this 
pulse is about 10\% higher in peak power than
the pulse shown in 
the main text Fig.~\ref{fig:AF}{\bf a}.
This is expected, since 
fixed-detuning pulses lack the 
additional degrees of freedom that are associated with 
being able to modulate the detuning. 
However, in analogy to 
the main text Fig.~\ref{fig:AF}{\bf a},
we see that the pulse 
tends to be relatively flat 
in amplitude, a feature we observed in all power-optimized 
pulses we generated in the course of numerous simulations. 
 
There are several reasons why 
step pulses should be 
replaced with 
AMFM pulses. 
In our opinion the two leading reasons are 
(i) amplitude-, frequency-, and phase-modulated pulses have lower power requirement as seen 
when comparing 
the main text Fig.~\ref{fig:AF}{\bf a}
and \fig{figstep} and  
(ii) in contrast to the sharp transitions 
in power levels characteristic for step pulses, 
amplitude-, frequency-, and phase-modulated pulses have a smooth pulse envelope, which 
eliminates ringing and the Gibbs phenomenon 
\cite{Gibbs} that 
accompanies sudden changes in power levels. 
 
In contrast to the straightforward construction 
of our amplitude-, frequency-, and phase-modulated pulses, 
finding the optimal pulse for step pulses 
requires a search in the 4D parameter 
space consisting of the number of segments, 
$N_{\rm seg}$, the detuning $\mu_0$, the integer $J$, 
and the parity $(\pm)$ of the pulse. 
While $N_{\rm seg}$ is discrete, and we found that 
good convergence is already achieved with 
relatively few segments, 
in terms of parity there are only 
two cases to check, 
and, if $\tau$ is pre-specified 
to a certain value, say, 
$\tau=300\,\mu{\rm s}\pm 1\,\mu{\rm s}$, 
the range of $J$ that falls into this interval is not large, 
and, moreover, $J$ is discrete,  
searching for the optimal detuning $\mu_0$ 
requires considerable computational overhead that 
is avoided using our ``single-shot'' 
AMFM approach. 
 
Concluding this subsection, we can say that our new 
linear algorithm is certainly general enough to 
encompass the important 
class of step pulses. 
Thus, if such pulses are required to run, e.g., 
existing quantum computers with existing controller 
hardware which requires step pulses as input, 
our method can be used to generate these 
pulses efficiently and directly. 
\subsection{Efficient arbitrary simultaneously entangling gates}
\label{sec:EASE}

In this section, we show how to use our method in conjunction
with the Efficient Arbitrary Simultaneously Entangling (EASE) gate 
protocol detailed in \cite{EASE}. To see how this may be achieved,
the only thing that is required is to show that the equations to
be solved are isomorphic. In particular, the null-space condition
(\ref{PG4}) is of the same structure as Eq.~(2) of \cite{EASE}
and the degree-of-entanglement condition (\ref{PG13a}) is of the same 
structure as Eq.~(3) of \cite{EASE}, which fully specify the problem
of solving for the EASE-gate pulse shapes. The rest of the EASE-gate
protocol follows immediately. The resulting pulse shapes can 
implement up to $N(N-1)/2$ $\xxgate$ gates simultaneously in a short time
for a given power budget.

\subsection{Sensitivity of the degree of entanglement}
\label{sec:Sensitivity}

\begin{figure}
\includegraphics[scale=.75,angle=0]{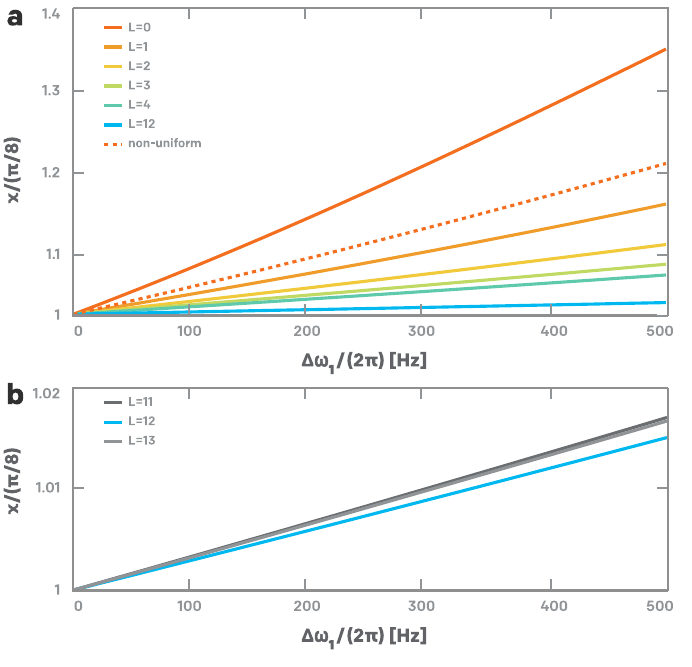}
\caption {\label{figchi}
Projection method: 
Sensitivity of the gate angle 
$\chi$ to drifts of the motional-mode frequencies 
$\omega_p$, $p=1,\ldots,P=N$, for $N=5$ ions. 
Shown is the normalized gate angle $|\chi|/(\pi/8)$ 
as a function of $\Delta\omega_1/(2\pi)$, 
the drift frequency of motional mode $p=1$. 
{\bf a.}
Orange solid curve: All five motional modes drift 
in unison from 0 to $2\pi\times 500\,$Hz. 
Orange dashed curve: The five motional 
modes drift independently as 
described in the text. The solid curves 
document the effect of active 
$\chi$ stabilization against motional-mode 
drifts as described in the text. 
{\bf b.}
Existence of a 
``sweet spot'' in the number of projected states. 
For the case chosen ($\tau=300\,\mu$s and 
5 ions) the sweet spot occurs for 12 projected 
states (blue curves in frames {\bf a} and 
{\bf b}). 
      }
\end{figure}
%
In this section we now explore 
the effects of motional-mode drifts 
on the gate angle $\chi$. For 
$N=5$ ions, two cases are investigated. (i) All modes 
$\omega_p$ drift in unison from 
$0\,$Hz to $+2\pi\times 500\,$Hz and 
(ii) individual modes drift 
independently. For case (ii), we simulated a 
case in which, chosen randomly, 
and with random signs of the drift direction, 
$\omega_1$ drifts from 
$0\,$Hz to $+2\pi\times 500\,$Hz, 
$\omega_2$ drifts from 
$0\,$Hz to $-2\pi\times 400\,$Hz, 
$\omega_3$ drifts from 
$0\,$Hz to $+2\pi\times 300\,$Hz,
$\omega_4$ drifts from 
$0\,$Hz to $-2\pi\times 500\,$Hz, and 
$\omega_5$ drifts from 
$0\,$Hz to $+2\pi\times 400\,$Hz. 
Figure~\ref{figchi} shows that 
although all drift amplitudes 
are substantially smaller than $1\,$kHz, 
the effect on the 
gate angle $\chi$ is substantial. 
 
The strong sensitivity of $\chi$ with 
respect to drifts in $\omega_p$ is due 
to the amplification effect of the 
relatively long pulse duration.
In order to compute $\chi$, we have to evaluate 
the double integral \eq{INTRO4}. 
Under the integral we have the term 
$\sin[\omega_p (t_2-t_1)]$, and
if we replace $\omega_p$ by $\omega_p + \Delta \omega_p$, then
the $\sin[\omega_p (t_2-t_1)]$ term becomes, in linear order,
$\sin[\omega_p (t_2-t_1)] + \cos[\omega_p(t_2-t_1)] 
\Delta \omega_p  (t_2-t_1)$.
Now, while $|\Delta \omega_p|$ is
at most $2\pi\times 500\,$Hz, which looks small,
and indicates that we might be able to
neglect the second term,
when we multiply the second term with $300\,\mu$s, 
which
is the maximum of $t_2-t_1$, we get 
$2\pi\times 0.0005\,{\rm MHz}\times 300\,\mu{\rm s} = 0.94$,
which is large. 
In fact, this term is so large that the 
linearization approximation breaks down. 
Therefore, the pulse length is the amplification mechanism 
and explains the strong sensitivity of $\chi$ 
to relatively small drifts in $\omega_p$. 
It also underpins the observed sensitivity 
(see Fig.~\ref{figchi}) 
with a detailed qualitative analytical understanding.
 
In order to counteract drifts in $\chi$, we suggest 
to monitor the value of $\chi$  
continuously and readjust the laser power that drives 
the $\xxgate$ gate 
if $\chi$ drifts away. This is a valid 
correction mechanism since the set of equations 
\eq{INTRO1} depends only on the {\it shape} 
of the pulse, but not on the pulse {\it amplitude}. 
Therefore, without compromising the 
validity of \eq{INTRO1}, the power 
can be continuously adjusted to keep 
$\chi$ within tolerable bounds. 
Of course, it may be difficult in practice 
to continuously monitor and readjust 
$\chi$. Nevertheless, at least in principle, 
this is a possible 
correction and stabilization mechanism. 
In analogy to our moments approach for 
active stabilization of the $\alpha$ 
conditions (\ref{eq:INTRO1}), 
it is also 
possible to encode 
active stabilization 
of $\chi$ in the pulse shape itself.
 
\subsection{Single-pulse 
active stabilization of the 
degree of entanglement: Projection method}
\label{sec:Proj}
Ideally, to actively 
stabilize $\chi_{ij}$ against $\omega_p$ 
fluctuations, integrated in the pulse-shape construction, 
we should require 
\begin{equation}
\label{eq:S14-1}
\chi_{ij,p}^{(k)} = 
\frac{\partial^k \chi_{ij}}{\partial \omega_p^k}=0,\ \ \ k=1,2,\ldots,K_{\chi},
\end{equation}
where $K_{\chi}$ is the maximal desired degree of $\chi$
stabilization.  
Since all pulse shapes, regardless of their maximal 
degree of stabilization $K_{\chi}$, need to 
satisfy both the decoupling conditions
(\ref{eq:INTRO1})  
between the motional modes and the computational states 
and the degree-of-entanglement 
condition (\ref{INTRO4a}) (where 
``$\pi/8$'' may be replaced by the actual desired degree of entanglement),
we may write
\begin{equation}
\label{eq:chiLambda}
\chi_{ij,p}^{(k)} = \vec{\Lambda}^T 
R_p^{(k)} \vec{\Lambda}=0,
\end{equation}
where
\begin{equation}
{R_{\alpha \beta ,p}^{(k)}} = (\vec{A}^{(\alpha)}){}^T S_p^{(k)} \vec{A}^{(\beta)}
\end{equation}
and
\begin{equation}
S_p^{(k)} = \frac{\partial^k S}{\partial 
\omega_p^k}. 
\end{equation}
To understand the consequences of 
(\ref{eq:S14-1}) [(\ref{eq:chiLambda}), 
respectively], we spectrally decompose $R_p^{(k)}$ according to 
\begin{equation}
R_p^{(k)} = \sum_{\nu=1}^{N_0} 
\lambda_{\nu,p}^{(k)} |\lambda_{\nu,p}^{(k)}\rangle \langle \lambda_{\nu,p}^{(k)} |,
\end{equation}
where $\lambda_{\nu,p}^{(k)}$ is the $\nu$-th 
eigenvalue of $R_p^{(k)}$ and $|\lambda_{\nu,p}^{(k)}\rangle$ is the 
corresponding eigenvector. 
Expanding $\vec \Lambda$ into the 
eigenstates of $R_p^{(k)}$, i.e., 
\begin{equation}
\vec{\Lambda} = \sum_{\nu=1}^{N_0} c_{\nu,p}^{(k)} |\lambda_{\nu,p}^{(k)}\rangle, 
\end{equation}
the stabilization condition (\ref{eq:chiLambda}) 
may then be written as
\begin{equation}
\label{eq:spec_decomp}
\chi_{ij,p}^{(k)} = \sum_{\nu=1}^{N_0} |c_{\nu,p}^{(k)}|^2 \lambda_{\nu,p}^{(k)} = 0. 
\end{equation}
Equation (\ref{eq:spec_decomp}) brings out 
the problem: 
The condition \eq{spec_decomp} can be satisfied only if not all of the eigenvalues 
$\lambda_{\nu,p}^{(k)}$ have the same sign. 
However, 
we can prove analytically (the proof 
is lengthy and not shown here) and confirmed 
numerically, that, for instance,    
$R_p^{(k=1)}$ is a definite matrix 
for all $p$, i.e., the eigenvalues of 
$R_p^{(k=1)}$ are all non-zero and have the same sign,
which makes it impossible to satisfy 
(\ref{eq:spec_decomp}) for $k=1$. 
We did, however, notice that 
only a few of the eigenvalues of 
$R_p^{(k=1)}$ are particularly large 
in absolute magnitude, which may be 
the ultimate reason for the strong 
sensitivity of $\chi_{ij}$ in linear 
order (see Fig.~\ref{figchi}). 
This observation suggests a   
strategy for actively stabilizing  
$\chi_{ij}$ against $\omega_p$-fluctuations: 
Projecting those components of 
the spectra of $R_p^{(k)}$ 
out of the null-space of $M$ 
(which can be assumed to already 
include stabilization of 
(\ref{eq:INTRO1}) against 
$\omega_p$-fluctuations)
that 
correspond to the eigenvalues with 
the largest absolute values. 
If we project out $L$ such components 
from each of the $R_p^{(k)}$ matrices, 
this leaves us with a null space of 
$N_0^{\prime}=N_0-P K_\chi L$ dimensions 
that now, to a large degree, actively 
stabilizes $\chi_{ij}$ against 
$\omega_p$-fluctuations. 
Following this projection step, we 
now use the techniques presented in 
\ref{PG}, applied to the reduced null space of 
$N_0^{\prime}$ dimensions, to satisfy the degree-of-entanglement 
condition with the smallest 
possible average power. 
 
To illustrate this technique, and 
focusing on the case of uniform drift 
of the motional modes from 0 to 
$500\,$Hz as defined above,   
we present 
in Fig.~\ref{figchi}{\bf a} the result of projecting 1,2,3,4, 
and 12 states from the null space that 
correspond to the eigenvalues with 
largest absolute values of 
$R_p^{(1)}$, $p=1,\ldots,5$. We see that 
already for a single projected state we 
achieve noticeable stabilization that improves 
further for 2, 3, and 4 projected states. 
This improvement continues if more states 
are projected, reaching an 
optimum (``sweet spot'') for 
12 projected states. This is illustrated 
in Fig.~\ref{figchi}{\bf b}, which 
shows the normalized $\chi$ for 11, 12, 
and 13 projected states. Therefore, while we found 
that projecting relatively few states 
always results in improved active 
stabilization, ``over-projection'' should 
be avoided, since it is both costly in power 
and does not improve $\chi$ stabilization 
any further. In fact, as expected, 
active $\chi$ stabilization, in analogy 
with stabilizing $\alpha$, requires increased 
levels of power. For example, for the case 
shown in Fig.~\ref{figchi}, the 
projection of 1, 2, 3, 4, and 12 states 
requires power levels of 1.5, 1.7, 2.0, 
2.1, and 3.3 with respect to the power 
level without projection. But we also see 
that projection is relatively inexpensive 
compared with the significant amount of 
stabilization gained. 
 
The projection technique works for all 
orders $k\ge 1$ of $R_p^{(k)}$. 
However, not much is gained 
by continuing the projection beyond 
$k=1$. The reason is the following. 
In our example, the best result, 
obtained by projecting the first 12 
states, brings down the variation 
in the relative $\chi$  
from 35\% to just 1.5\%.
These 1.5\%, however, are mostly due to 
the residual slope of the first-order 
stabilization, so that second-order 
stabilization would not contribute much, 
other than computational effort and 
power expended. We see this in 
the following way. The slope of 
the first-order stabilization for 
12 projected states at 0 motional-mode 
drift is $3\times 10^{-5}\,$/Hz. 
Therefore, at $500\,$Hz motional-mode drift, 
the variation in the relative value of $\chi$
is 0.015, i.e. 1.5\%. This is exactly the 
amount we read off in  
Fig.~\ref{figchi}{\bf b} for the case 
of 12 projected states. 
Therefore, the residual variation is 
mostly due to the first order, and 
stabilizing the second order will 
have a negligible effect. 
Nevertheless, as shown in 
Fig.~\ref{figchi}{\bf a}, and given 
an unstabilized variation of 
35\%, 
active first-order stabilization 
via projection, 
which brings this variation down to 
about 1.5\% (see Fig.~\ref{figchi}{\bf b}), is 
already significant for the stabilization 
of quantum-computer operation. 
While, as discussed above, it is not possible 
to implement the moments strategy 
(\ref{eq:S14-1}) with a single pulse, 
working with two different pulses, 
each directed 
at a different one of the two ions 
participating in the gate, 
(\ref{eq:S14-1}) can in fact be realized 
and results in the moments methods 
described in the following two sections. 
 
\subsection{Two-pulse active stabilization 
of the degree of entanglement: 
Moments method}
\label{SOM-ASMM} 
\begin{figure}
\centering
\includegraphics[scale=0.8,angle=0]{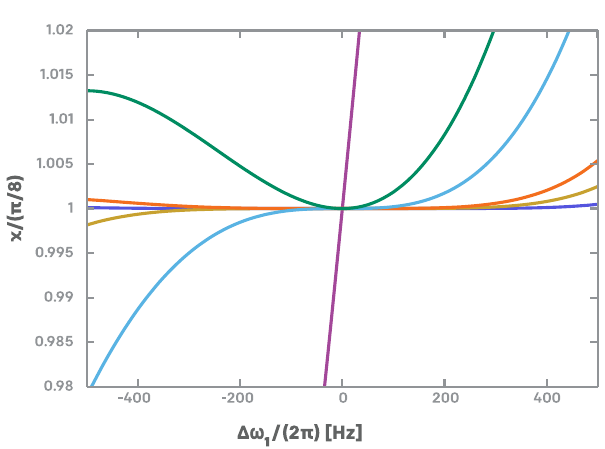}
\caption {\label{FigS3}
Two-pulse moments approach: 
Sensitivity of the gate angle 
$\chi$ to uniform drifts of the 
motional-mode frequencies 
$\omega_p$, $p=1,\ldots,P=N$ for 
$N=5$ ions as a function of 
the drift frequency $\Delta\omega_1$ 
of the first motional mode 
for the first six stabilization 
orders $K_{\chi}$. Different 
color lines correspond to 
different stabilization orders 
$K_{\chi}$. 
Purple: $K_{\chi}=0$; 
green: $K_{\chi}=1$;
cyan: $K_{\chi}=2$;
orange: $K_{\chi}=3$;
yellow: $K_{\chi}=4$;
blue: $K_{\chi}=5$. 
Compared with Fig.~\ref{figchi}, 
the two-pulse moments 
method provides significantly 
improved $\chi$ stabilization. 
      }
\end{figure}
%

While it is impossible to satisfy 
(\ref{eq:S14-1}) using 
identical pulses directed at both 
ions $i$ and $j$, the condition 
(\ref{eq:S14-1}) can be satisfied using 
different pulses directed at ions 
$i$ and $j$. 
Let $g^{(i)}(t)$ and $g^{(j)}(t)$ be the pulse 
functions directed at ions $i$ and $j$, 
respectively, and let 
$\hat{\vec F}$ and $\hat{\vec G}$ 
be the null-space expansion 
amplitudes of 
$g^{(i)}(t)$ and $g^{(j)}(t)$, 
respectively. Then, 
condition \eq{chi} in the main text implies 
\begin{equation}
\chi = \frac{\pi}{8} = 
\hat{\vec F}^T R\, \hat{\vec G}, 
\label{new-S15-1}
\end{equation}
which has to be solved under the 
conditions [see (\ref{eq:S14-1})] 
\begin{equation}
\vec F^T R_p^{(k)} \vec G = 0, \ \ \ 
p=1,\ldots,P,\ \ \ 
k=1,\ldots,K_{\chi}, 
\label{new-S15-2}
\end{equation}
where $\vec F$ and $\vec G$ 
are any un-normalized versions of 
$\hat{\vec F}$ and $\hat{\vec G}$. 
Unlike in the single-pulse case   
discussed 
in \ref{sec:Proj},
where, due to the 
observed definiteness 
of $R_p^{(1)}$, it 
was 
impossible to 
satisfy (\ref{eq:S14-1}), 
working with two 
different pulses, 
$\vec F$ and $\vec G$, 
(\ref{new-S15-2}) can be satisfied 
as soon as 
$\vec F$ and $\vec G$ are 
orthogonal to each other with 
respect to $R_p^{(k)}$. 
 
An explicit solution of 
(\ref{new-S15-1}) and 
(\ref{new-S15-2}) can be constructed 
in the following way. We start 
by choosing $\vec G$ to be 
the power-optimal pulse as computed 
in the single-pulse case described 
in \ref{PG}. Then, we construct the 
space ${\cal P}$, which is 
spanned by the vectors 
$\vec v_p^{(k)}=R_p^{(k)}\vec G$, 
$p=1,\ldots,P$, $k=1,\ldots,K_{\chi}$. We 
also define the space 
${\cal Q}$, which is the orthogonal 
complement of ${\cal P}$ with 
respect to the null space. With 
these definitions, taking 
$\vec F$ out of ${\cal Q}$, 
we obtain the most 
power-optimal solution for $\vec F$ 
by choosing 
$\vec F = \hat Q \vec G$, 
where $\hat Q$ projects into 
the ${\cal Q}$ space. At this point 
the normalizations of 
$\vec F$ and $\vec G$ 
are still two free parameters. 
We use the 
first scaling freedom to obtain 
$|\hat{\vec F}|=|\hat{\vec G}|$ 
(symmetric pulse power) and use 
the second scaling freedom to satisfy 
(\ref{new-S15-1}). 
\begin{figure*}
\centering
\includegraphics[width=\textwidth,angle=0]{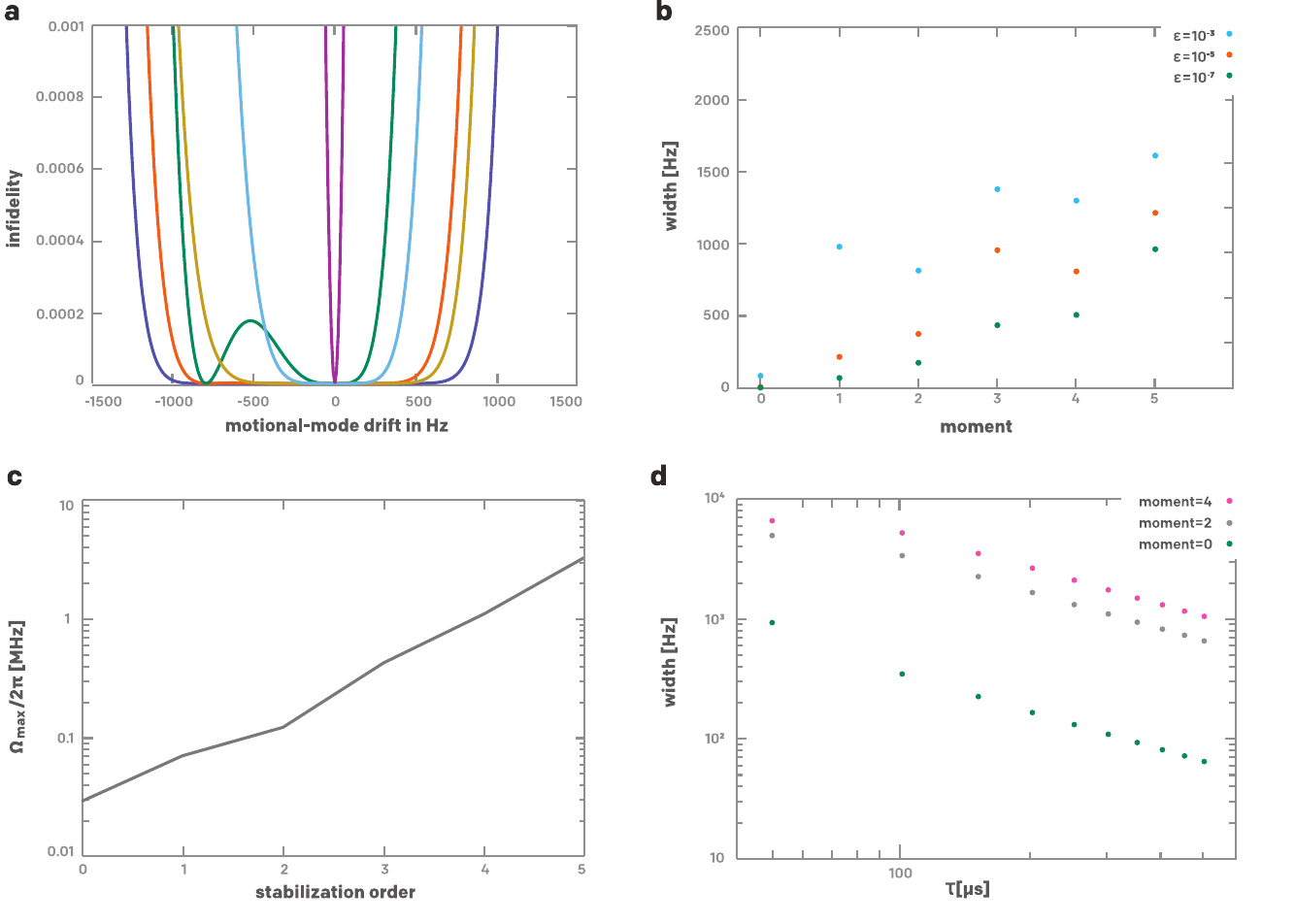}
\caption{Stabilization of the control pulses. 
{\bf a}. Infidelity (see SI section \ref{Stab} for detail) 
as a function of the motional-mode frequency drift $\Delta f$. 
All mode frequencies were drifted according to 
$\omega_p \mapsto \omega_p + 2\pi \Delta f$. 
{\bf b}. The width of the infidelity curves in {\bf a} for 
various error tolerances 
$\epsilon = 10^{-3}, 10^{-5}, \text{ and } 10^{-7}$, 
as a function of the highest moment $K_{\chi}$ 
of stabilization. 
{\bf c}. The maximal power requirement $\max_t|g_K(t)|$ 
of the control pulses  
as a function of the highest moment $K_{\chi}$ of stabilization. 
The power requirement 
suggests an exponential 
scaling 
of power
in the 
order
of 
stabilization 
$K_{\chi}$. 
{\bf d}. Width of the infidelity curves for various different 
orders of stabilization $K_{\chi}=0, 2,\text{ and } 4$, as a 
function of the gate duration $\tau$ for a fixed 
error tolerance level $\epsilon = 10^{-3}$. 
The data suggests $\sim 1/\tau$ scaling of the width.}
\label{fig:chi-stabilize}
\end{figure*}

For the same case of uni-directional 
$\omega_p$ drift as used in 
Fig.~\ref{figchi}, 
the performance of active 
two-pulse $\chi$ stabilization is 
illustrated in Fig.~\ref{FigS3} 
and can be compared with 
the performance 
of the single-pulse projection 
method illustrated in Fig.~\ref{figchi}. 
We see that while the projection method 
at best cuts the relative $\chi$ drift error 
down to 1.5\% at $500\,$Hz drift, the 
relative error in the case of 
$K_{\chi}=5$ stabilization 
orders is smaller than 
1 permille even at $500\,$Hz drift. 
In addition, while in the 
projection method the stabilization is 
of linear order around zero motional-mode 
drift, the moments method produces 
favorable non-linear stabilization of 
order $K_\chi+1$ around 
zero motional-mode drift 
(see Fig.~\ref{FigS3}). 
The power of the moments method is best 
appreciated in Fig.~\ref{FigS3} 
by comparing the behavior of 
the unstabilized $\chi$ 
(the nearly vertical, purple line in 
Fig.~\ref{FigS3} around 
zero motional-mode drift; $K_{\chi}=0$) with 
the shape of $\chi$ as a function 
of motional-mode drift for different 
$K_{\chi}\geq 1$. 
Even for small $K_{\chi}\geq 1$, substantial
stabilization is observed. 
 
In Fig.~\ref{fig:chi-stabilize} we show 
a summary of various aspects of 
the two-pulse active $\chi$ stabilization 
method. Figure~\ref{fig:chi-stabilize}{\bf a} 
shows the infidelity of the two-pulse 
active-stabilization method 
as a function of mode-frequency drift. 
We see that infidelities $\ll 10^{-3}$ 
can be achieved for motional-mode 
frequency drifts in the range 
$\pm 200\,$Hz for all $K_{\chi}\geq 1$ 
and and an infidelity of $\ll 10^{-4}$ 
can be achieved over a motional-mode 
frequency drift in the range 
$\pm 500\,$Hz for $K_{\chi}=5$. 
Figure~\ref{fig:chi-stabilize}{\bf b} 
shows 
a summary of 
frequency widths of $\chi$ 
stabilization
for infidelity cut-offs of 
$10^{-3}$, $10^{-5}$, and 
$10^{-7}$, respectively. 
On average we observe a linear 
increase of frequency width 
with the $\chi$ stabilization order. 
While these results are promising, 
they do come with a price. 
Figure~\ref{fig:chi-stabilize}{\bf c} 
shows the power requirement 
as a function of stabilization 
order $K_{\chi}$. 
In constrast with the 
linear power scaling of 
$\alpha$ stabilization shown 
in Fig.~\ref{fig:stabilize}{\bf c}, 
Fig.~\ref{fig:chi-stabilize}{\bf c}  
indicates that the required 
pulse power 
in the case of $\chi$ 
stabilization increases 
exponentially with the 
stabilization order. 
This, however, should not 
discourage us from using 
the moments method for active 
two-pulse stabilization of 
$\chi$, since, as shown in 
Figs.~\ref{FigS3} 
and \ref{fig:chi-stabilize}, 
significant $\chi$ 
stabilization 
is already achieved for  
relatively small $K_{\chi}$.
Figure~\ref{fig:chi-stabilize}{\bf d}
shows the scaling of frequency 
width for several stabilization 
orders $K_{\chi}$ as a function 
of gate duration $\tau$. 
Similar to the 
results shown in 
Fig.~\ref{fig:stabilize}{\bf d}, 
we observe that the frequency 
width is inversely proportional 
to the gate duration.

\subsection{Two-pulse active stabilization 
of the degree of entanglement:  
Hybrid method}
\label{SOM-ASHM} 
\begin{figure}
\includegraphics[scale=.8,angle=0]{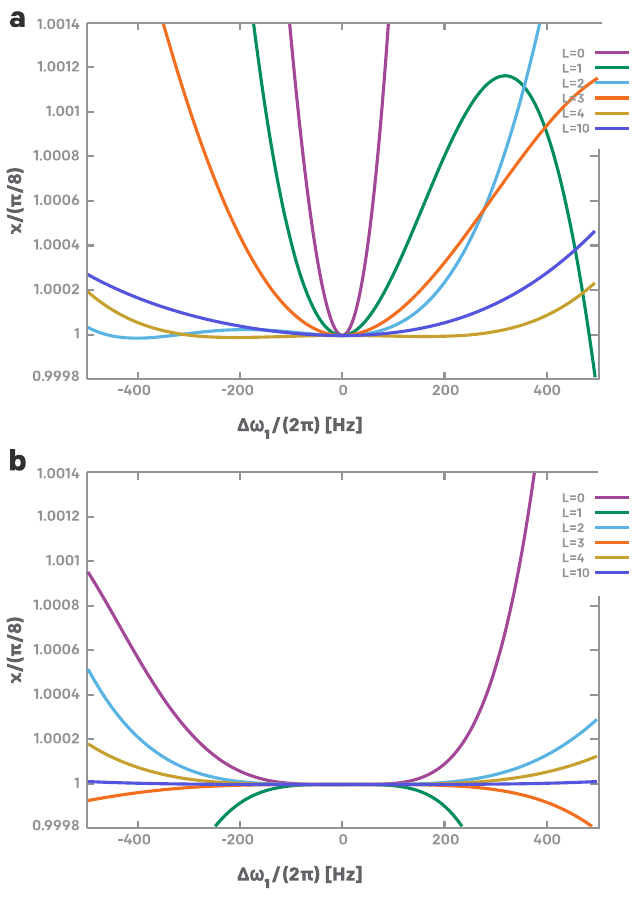}
\caption {\label{fighybrid}
Performance of the hybrid method. 
{\bf a}. For $K_{\chi}=1$,  
$L=0,1,2,3,4$, and $L=10$ vectors 
with the largest absolute value 
of their corresponding eigenvalues 
are projected out. 
{\bf b}. Same as {\bf a}. but with 
$K_{\chi}=3$. {\bf b}. shows that 
compared with {\bf a}. larger 
$K_{\chi}$ results in a broader 
stabilization region, which is 
widened even further by projection.
      }
\end{figure}

Combining the most advantageous 
features of both the moments and 
projection methods, we arrive 
at the hybrid method. 
In this method, for desired 
stabilization order $K_{\chi}$,
we first construct 
the pulse $\hat{\vec G}$ as described in 
section \ref{sec:Proj}, i.e., as a single pulse 
in a null space of dimension 
$N_0'=N_0-P L$, 
where 
we projected out those $L$ eigenvectors 
from each matrix $R_p^{(1)}$, 
$p=1,\ldots,P$, 
that correspond to the $L$ 
eigenvalues with the 
largest absolute 
values. We then proceed 
with the construction of 
the pulse $\hat{\vec F}$
as described in section \ref{SOM-ASMM}. 
Thus, 
even before construction of 
the pulse $\hat{\vec F}$, 
we eliminate those null-space 
components from the 
pulse $\hat{\vec G}$ that 
potentially produce the 
most sensitivity in $\chi$. 
As documented in Fig.~\ref{fighybrid}, 
and compared with the single-pulse 
projection method 
(see \ref{sec:Proj}) and the bare 
two-pulse moments method 
(see \ref{SOM-ASMM}), 
the hybrid method yields the best 
results in terms of active 
$\chi$ stabilization. 
 
Figure~\ref{fighybrid}{\bf a}, 
for the same case of uniform 
mode-frequency drift as 
used in Figs.~\ref{figchi}, 
\ref{FigS3}, and \ref{fig:chi-stabilize}, 
shows the result of 
$\chi$ stabilization 
for $K_{\chi}=1$ and 
$L$ vectors with the 
largest absolute values
of their 
eigenvalues 
projected out. 
We see that already 
for the case $K_{\chi}=1$, 
compared with  
the bare moment method for 
$K_{\chi}=1$ (see 
Fig.~\ref{FigS3}), the gain 
in stabilization is substantial. 
This observation 
is important since, 
because of the 
exponential power 
cost of the two-pulse 
method in 
$K_{\chi}$, 
it is advantageous, 
for a given 
stabilization target, to 
stabilize with the smallest possible 
$K_{\chi}$. 
Figure~\ref{fighybrid}{\bf b} shows 
that with a slightly larger
$K_{\chi}$ [$K_{\chi}=3$ in 
Fig.~\ref{fighybrid}{\bf b}], 
a substantial broadening of 
the stabilization region can 
be achieved. 
Figure~\ref{fighybrid}{\bf b}
also shows that by projecting out 
$L=10$ vectors, $\chi$ stabilization 
on the level of $10^{-5}$ can 
be achieved for $500\,$Hz 
motional-mode drift. 
%
%

\subsection{Broadband sequence}
\label{sec:Broad}

A host of compensation pulse sequences that mitigate the errors in the single-qubit gate are known (see \cite{Broadband} and the references therein) and similar techniques can indeed be used to mitigate the errors in $\chi_{ij}$ that arise from, e.g., relative offsets in $\eta_i^p$ or $g(t)$. We refer interested readers to \cite{Ivanov,Murphy}, where a variety of broadband behaviors have been explored. Below, we show an example for completeness.

Consider a broadband behavior of the Solovay-Kitaev (SK) sequence in \cite{Broadband} to compensate for the inexact $\chi_{ij}$ up to first order. Higher orders or other compensation techniques, such as those that rely on Suzuki-Trotter sequences, may straightforwardly be employed. Typically, the SK compensation sequence is discussed in the context of single-qubit operators, where, for small error strength $\epsilon$,
\small
\begin{align}
&R(\theta,0) - R(2\pi(1+\epsilon),-\phi_\text{SK}) R(2\pi(1+\epsilon),\phi_\text{SK}) R(\theta(1+\epsilon),0)  \nonumber \\
&= O(\epsilon^2),
\end{align}
\normalsize
where 
\begin{equation}
R(\theta,\phi) =  \exp\{-i\theta[\cos(\phi) \sigma_x + \sin(\phi) \sigma_y]/2\}
\end{equation}
and
\begin{equation}
\phi_\text{SK} = \cos^{-1}(-\theta/4\pi).
\end{equation}
A straightforward extension to the two-qubit $\xxgate(\theta)$ gate may be done to result in
\begin{equation}
\xxgate(\theta) - \xmphigate(2\pi(1+\epsilon)) \xphigate(2\pi(1+\epsilon)) \xxgate(\theta(1+\epsilon)) = O(\epsilon^2),
\end{equation}
where
\begin{align}
\xphigate(\theta) = (1\otimes \rzgate(\phi_\text{SK})) \xxgate(\theta) (1\otimes \rzgate(-\phi_\text{SK})), \nonumber \\
\xmphigate(\theta) = (1\otimes \rzgate(-\phi_\text{SK})) \xxgate(\theta) (1\otimes \rzgate(\phi_\text{SK})),
\end{align}
and 
\begin{equation}
\rzgate(\phi) = \exp(-i\theta\sigma_z/2).
\end{equation}
The choice of application of $\rzgate$ gates on the second qubit is arbitrary, and can indeed instead be performed on the first qubit without loss of generality. Because the errors in $\chi_{ij}$ incur in one well-defined direction of $\sigma_x\sigma_x$ in the 15-dimensional hyper-Bloch sphere, the single-qubit compensation-pulse techniques become straightforwardly applicable.

\subsection{Direct implementation 
of Fourier-basis pulse function} 
\label{sec:Direct}

According to \cite{AM}, in the Lamb-Dicke regime, the interaction Hamiltonian for the ion-chain system, subjected to a dual-tone, symmetric blue- and red-sideband beam with detuning $\pm \mu$, in the $x$ basis is
\small
\begin{equation}
H_\text{dual-tone}(t)=\sum_{i=1}^N \sum_{p=1}^P \Omega_i (t) \eta_p^i \sin(\mu t) (a_p e^{-i\omega_p t} + a_p^\dagger e^{i\omega_p t}) \sigma_x^i,
\end{equation}
\normalsize
where $a_p$ and $a_p^\dagger$ are the annihilation and creation operators of the $p$th motional mode, respectively.
Consider now a multi-tone beam with amplitudes $\Omega_{i,n}(t)$ and detuning frequencies $\pm \mu_n$, where $n=1,2,..,N_A$. This results in the Hamiltonian
\begin{align}
\label{eq:multitone}
H_\text{multi-tone}(t)=\sum_{i=1}^N \sum_{p=1}^P \sum_{n=1}^{N_A} &\Omega_{i,n} (t) \eta_p^i \sin(\mu_n t) \nonumber \\ 
&(a_p e^{-i\omega_p t} + a_p^\dagger e^{i\omega_p t}) \sigma_x^i.
\end{align}
Define
\begin{equation}
\label{eq:fdef}
f_{ip}(t) = \eta_p^i g_i(t),
\end{equation}
where
\begin{equation}
\label{eq:gdef}
g_i(t) = \sum_{n=1}^{N_A} 
\Omega_{i,n} (t)\sin(\mu_n t), 
\end{equation}
where $N_A$ is chosen sufficiently large 
to achieve convergence. 
Inserting \eq{fdef} in \eq{multitone}, we obtain
\begin{equation}
\label{eq:multitone2}
H_\text{multi-tone}(t)=\sum_{i=1}^N \sum_{p=1}^P f_{ip}(t) (a_p e^{-i\omega_p t} + a_p^\dagger e^{i\omega_p t}) \sigma_x^i,
\end{equation}
which induces the system evolution over the gate time $\tau$ described by
\begin{align}
\label{eq:multievol}
U_\text{multi-tone}(t) \nonumber \\
= 
\exp\bigg\{&-i\int_0^\tau\ dt H_\text{multi-tone}(t) -\frac{1}{2}\int_0^\tau dt_2 \int_0^{t_2} dt_1 \nonumber \\
& [H_\text{multi-tone}(t_2),H_\text{multi-tone}(t_1)] \bigg\},
\end{align}
as shown in \cite{AM} using Magnus' formula. Inserting \eq{multitone2} in \eq{multievol}, together with \eq{fdef} and \eq{gdef}, we obtain, up to a global phase,
\begin{align}
U_\text{multi-tone}(t) \nonumber \\
= \exp\Bigg\{&-i\left[\sum_{i=1}^N \sum_{p=1}^P \left(\alpha_{ip} \eta_p^i a_p + \alpha^*_{ip} \eta_p^i a_p^\dagger \right) \sigma_x^i \right] \nonumber \\
&+ i \sum_{i,j=1; i\neq j}^N \chi_{ij} \sigma_x^i \sigma_x^j \Bigg\},
\end{align}
where
\begin{equation}
\label{eq:alpha_g}
\alpha_{ip} = \int_0^\tau g_i(t) e^{-i\omega_p t} dt,
\end{equation}
$\alpha^*_{ip}$ denotes its complex conjugate,
and
\begin{equation}
\label{eq:alpha_chi}
\chi_{ij} = \sum_{p=1}^P \eta_p^i \eta_p^j \int_0^\tau dt_2 \int_0^{t_2} dt_1 g_i(t_2) g_j(t_1) \sin[\omega_p(t_2-t_1)].
\end{equation}
Comparing \eq{alpha_g} and \eq{alpha_chi} 
with 
\eq{alpha} and \eq{chi}, 
respectively, together with \eq{gdef}, and 
assuming  
$g_i(t)=g_j(t)=g(t)=\sum_{n=1}^{N_A} 
A_n \sin(2\pi n t/\tau)$, 
as we did in the main text, 
we see that $\Omega_{i,n}(t) = A_n$ and $\mu_n = 2\pi n/\tau$ implements the pulse function that implements the desired $\xxgate$ gate. 


\begin{thebibliography}{99}
 
\bibitem{ar:HHL}
A. W. Harrow, A. Hassidim, S. Llyod,
Quantum algorithm for solving linear systems of equations.
Phys. Rev. Lett. {\bf 15}, 150502 (2009).

\bibitem{ar:GenMod}
M. Benedetti, D. Garcia-Pintos, O. Perdomo, V. Leyton-Ortega, Y. Nam, A. Perdomo-Ortiz,
A generative modeling approach for benchmarking and training shallow quantum circuits.
{\it npj Quant. Inf.} {\bf 5}, 45 (2019).

\bibitem{ar:Shor}
P. W. Shor, 
Polynomial-time algorithms for prime factorization and discrete logarithms on a quantum computer. 
{\it SIAM Rev.} {\bf 41}, 303--332 (1999).

\bibitem{ar:NF}
M. Reiher, N. Wiebe, K. M. Svore, D. Wecker, M. Troyer,
Elucidating reaction mechanisms on quantum computers.
{\it Proc. Natl. Acad. Sci. U.S.A.} {\bf 114}, 7555--7560 (2017).

\bibitem{ar:HINT}
Y. Nam and D. Maslov,
Low cost quantum circuits for classically intractable instances of the Hamiltonian dynamics simulation problem.
{\it npj Quant. Inf.} {\bf 5}, 44 (2019).

\bibitem{finance1}
S. Lloyd, M. Mohseni, P. Rebentrost,
Quantum principal component analysis.
{\it Nat. Phys.} {\bf 10}, 631 (2014).

\bibitem{finance2}
R. Or\'us, S. Mugel, E. Lizaso,
Quantum computing for finance: overview and prospects.
{\it Rev. in Phys.} {\bf 4}, 100028 (2019).

\bibitem{IonQ}
K. Wright {\it et al.},
Benchmarking an 11-qubit quantum computer.
{\it Nat. Commun.} {\bf 10}, 5464 (2019).

\bibitem{UMD}
S.~Debnath, N.~M. Linke, C. Figgatt, K.~A. Landsman, 
K. Wright, C. Monroe,
Demonstration of a small programmable quantum
computer with atomic qubits.
{\it Nature} {\bf 536}, 63--66 (2016).

\bibitem{Solidstate}
IBM Research. 
Quantum Experience. 
\href{http://www.research.ibm.com/quantum/}{http://www.research.ibm.com/quantum/},
Accessed April 27, (2019);
Rigetti Computing.
Quantum Cloud Services.
\href{https://www.rigetti.com/qcs}{https://www.rigetti.com/qcs},
Accessed April 27, (2019).

\bibitem{MS-1} 
K. M{\o}lmer, A. S{\o}rensen, 
Multiparticle Entanglement of Hot Trapped Ions, 
{\it Phys. Rev. Lett.} {\bf 82}, 1835-1838 (1999). 

\bibitem{AM}
S.-L. Zhu, C. Monroe, L.-M. Duan,
Arbitrary-speed quantum gates within large ion crystals through minimum control of laser beams.
{\it Europhys. Lett.} {\bf 73}, 485 (2006).

\bibitem{AM2} T. Choi, S. Debnath, T. A. Manning, 
C. Figgatt, Z.-X. Gong, L.-M. Duan, and C. Monroe, 
Optimal Quantum Control of 
Multimode Couplings between 
Trapped Ion Qubits for Scalable Entanglement.
{\it Phys. Rev. Lett.} {\bf 112}, 190502 (2014). 

\bibitem{Gaebler2016}
 J. P. Gaebler, T. R. Tan, Y. Lin, Y. Wan, R. Bowler, A. C. Keith, S. Glancy, K. Coakley, E. Knill, D. Leibfried, D. J. Wineland,
 High-fidelity universal gate set for ${^{9}\mathrm{Be}}^{+}$ ion qubits. 
 {\it Phys. Rev. Lett.} {\bf 117}, 060505 (2016).

\bibitem{Ballance2016}
 C. J. Ballance, T. P. Harty, 
 N. M. Linke, M. A. Sepiol, D. M. Lucas, 
 High-fidelity quantum logic gates 
 using trapped-ion hyperfine qubits. 
 {\it Phys. Rev. Lett.} {\bf 117}, 060504 (2016).

\bibitem{Harty2014}
T. P. Harty, D. T. C. Allcock, C. J. Ballance, L. Guidoni, H. A. Janacek, N. M. Linke, D. N. Stacey, D. M. Lucas, 
High-fidelity preparation, gates, memory, and readout of a trapped-ion quantum bit. 
{\it Phys. Rev. Lett.} {\bf 113}, 220501 (2014).

\bibitem{FM}
P.~H. Leung, K.~A. Landsman, C. Figgatt, N.~M. Linke, C. Monroe, K.~R. Brown,
Robust 2-qubit gates in a linear ion crystal using a frequency-modulated driving force.
{\it Phys. Rev. Lett.} {\bf 120}, 020501 (2018).

\bibitem{PM}
T.~J. Green, M.~J. Biercuk, 
Phase-modulated decoupling 
and error suppression in qubit-oscillator systems.
{\it Phys. Rev. Lett.} {\bf 114}, 120502 (2015).

\bibitem{ar:VQE}
Y. Nam~{\it et al.}, 
Ground-state energy estimation of the water molecule on a trapped ion quantum computer.
{\it npj Quantum Inf.} {\bf 6}, 33 (2020).

\bibitem{ar:QAOA}
G. E. Crooks,
Performance of the quantum approximate optimization algorithm on the maximum cut problem.
\href{https://arxiv.org/abs/1811.08419}{https://arxiv.org/abs/1811.08419} (2018).

\bibitem{ar:QFT}
Y. Nam, Y. Su, D. Maslov,
Approximate quantum Fourier transform with $O(n\log(n))$ {T} gates.
{\it npj Quantum Inf.} {\bf 6}, 26 (2020). 

\bibitem{ar:Kutin}
T. G. Draper, S. A. Kutin, E. M. Rains, K. M. Svore,
A logarithmic-depth quantum carry-lookahead adder.
{\it Quant. Inf. Comp.} {\bf 6}, 351--369 (2006).

\bibitem{ar:Babbush}
R. Babbush~{\it et al.},
Encoding electronic spectra in quantum circuits with linear {T} complexity.
{\it Phys. Rev. X} {\bf 8}, 041015 (2018).

\bibitem{ar:Grover}
L. K. Grover, 
Quantum mechanics helps in searching for a needle in a haystack.
{\it Phys. Rev. Lett.} {\bf 79}, 325 (1997).
 
\bibitem{ar:kSAT}
F. A. Aloul, A. Ramani, I. L. Markov, K. A. Sakallah,
Solving difficult SAT instances in the presence of symmetry.
{\it Proc. Des. Automat. Conf.} 731--736 (2002).

\bibitem{ar:BH}
S. Bravyi, J. Haah,
Magic state distillation with low overhead.
{\it Phys. Rev. A} {\bf 86}, 052329 (2012).

\bibitem{ar:Campbell}
J. O'Gorman, E. T. Campbell,
Quantum computation with realistic magic state factories.
{\it Phys. Rev. A} {\bf 95}, 032338 (2017).

\bibitem{WEIZ-PRL} 
Y. Shapira, R. Shaniv, 
T. Manovitz, N. Akerman, R. Ozeri, 
Robust Entanglement Gates for Trapped-Ion Qubits,
Phys. Rev. Lett. {\bf 121} 180502 (2018). 

\bibitem{EASE}
N. Grzesiak, R. Bl\"umel, K. Beck, K. Wright, V. Chaplin, J. Amini, N. Pisenti, S. Debnath, J.-S. Chen, Y. Nam,
Efficient Arbitrary Simultaneously Entangling Gates on a trapped-ion quantum computer.
\href{https://arxiv.org/abs/1905.09294}{https://arxiv.org/abs/1905.09294} (2019); 
{\it Nat. Commun.} {\bf 11:2963} (2020), 
\href{https://doi.org/10.1038/s41467-020-16790-9}{https://doi.org/10.1038/s41467-020-16790-9}

\bibitem{GREENPAP} Y. Wu, S.-T. Wang, L.-M. Duan, 
Noise Analysis for High-Fidelity Quantum Entangling Gates in an Anharmonic Linear Paul Trap.
{\it Phys. Rev. A} {\bf 97}, 062325 (2018). 

\bibitem{GR}
J. J. Garc\'{i}a-Ripoll, P. Zoller, J. I. Cirac,
Coherent control of trapped ions using off-resonant lasers.
{\it Phys. Rev. A} {\bf 71}, 062309 (2005).

\bibitem{PARA1}
C. Figgatt, A. Ostrander, N.~M. Linke, K.~ A. Landsman, D. Zhu, D. Maslov, C. Monroe,
Parallel entangling operations on a universal ion trap quantum computer.
{\it Nature} {\bf 572}, 368--372 (2019).

\bibitem{PARA2}
Y. Lu, S. Zhang, K. Zhang, W. Chen, Y. Shen, J. Zhang, J.-N. Zhang, K. Kim,
Global entangling gates on arbitrary ion qubits.
{\it Nature} {\bf 572}, 363-–367 (2019).
 
\bibitem{NUM-REC} 
W. H. Press, S. A. Teukolsky, 
W. T. Vetterling, B. P. 
Flannery, 
Numerical Recipes, 
second edition 
(Cambridge University Press, 
Cambridge, 1992). 

\bibitem{OXFORD-GATE}
V. M. Sch\"afer, C. J. Ballance, 
K. Thirumalai, L. J. Stephenson, 
T. G. Ballance, A. M. Steane, D. M. Lucas,  
Fast quantum logic gates 
with trapped-ion qubits,
Nature {\bf 555}, 75--78 (2018).
 
%
\bibitem{UMDPRL} 
R. Bl\"umel, 
N. Grzesiak, 
N. H. Nguyen, 
A. M. Green, 
M. Li, 
A. Maksymov, 
N. M. Linke,  
Y. Nam, 
Efficient Stabilized Two-Qubit Gates on a Trapped-Ion Quantum Computer. 
Phys. Rev. Lett. 
{\bf 126}, 
220503 (2021).

\bibitem{Sideband}
J.-S. Chen, K. Wright, N. C. Pisenti, D. Murphy, K. M. Beck, K. Landsman, J. M. Amini, Y. Nam,
Efficient sideband cooling protocol for long trapped-ion chains.
\href{https://arxiv.org/abs/2002.04133}{https://arxiv.org/abs/2002.04133} (2020).

\bibitem{HONEYWELL} 
\href{https://www.honeywell.com/en-us/company/quantum}
{https://www.honeywell.com/en-us/company/quantum} 
(Accessed September 13, 2020).

\bibitem{PHYS-TODAY}
H. Ball, M. J. Biercuk, and 
M. R. Hush, 
Quantum firmware and the 
quantum computing stack, 
Physics Today, March 2021, 29--34.

\bibitem{SL-PRL} 
A. E. Webb, S. C. Webster, S. Collingbourne, 
D. Bretaud, A. M. Lawrence, S. Weidt, F. Mintert, 
W. K. Hensinger, 
Resilient Entangling Gates for Trapped Ions, 
Phys. Rev. Lett. {\bf 121} 180501 (2018).


\bibitem{ar:GW}
M. X. Goemans, 	D. P. Williamson,
Improved approximation algorithms for maximum cut and satisfiability problems using semidefinite programming.
{\it J. ACM} {\bf 42}, 1115-1145 (1995).

\bibitem{ar:HeisenNN}
A. M. Childs, D. Maslov, Y. Nam, N. J Ross, Y. Su,
Toward the first quantum simulation with quantum speedup.
{\it Proc. Natl. Acad. Sci. U.S.A.} {\bf 115}, 9456--9461 (2018).

\bibitem{ar:Markov}
V.~V. Shende, I.~L. Markov, S.~S. Bullock,
Minimal universal two-qubit controlled-NOT-based circuits.
{\it Phys. Rev. A} {\bf 69}, 062321 (2004).

\bibitem{lee2020even}
J. Lee, D. Berry, C. Gidney, W. J. Huggins, J. R. McClean, N. Wiebe, R. Babbush,
Even more efficient quantum computations of chemistry through tensor hypercontraction.
\href{https://arxiv.org/abs/2011.03494}{https://arxiv.org/abs/2011.03494} (2020).

\bibitem{ar:reltof}
D. Maslov,
On the advantages of using relative phase Toffolis with an application to multiple control Toffoli optimization.
{\it Phys. Rev. A} {\bf 93}, 022311 (2016).

\bibitem{LD} D. J. Wineland, C. Monroe,
W. M. Itano, D. Leibfried,
B. E. King, D. M. Meekhof, 
Experimental Issues in Coherent Quantum-State 
Manipulation of Trapped Atomic Ions.
{\it J. Res. Natl. Inst. Stand. Technol.}
{\bf 103}, 259--328 (1998).

\bibitem{MMETA} 
C. Marquet, F. Schmidt-Kaler, D. F. V. James, 
Phonon-phonon interactions due to non-linear effects in a linear ion trap.
{\it Appl. Phys. B} {\bf 76}, 199--208 (2003). 

\bibitem{Gibbs} E. Hewitt, R. E. Hewitt,
The Gibbs-Wilbraham phenomenon: An episode in Fourier analysis, 
{\it Archive for History of Exact Sciences} {\bf 21}, 129--160 
(1979). 

\bibitem{Ivanov}
Svetoslav S. Ivanov, Nikolay V. Vitanov,
Composite two-qubit gates.
{\it Phys. Rev. A} {\bf 92}, 022333 (2015).

\bibitem{Murphy}
Daniel C. Murphy, Kenneth R. Brown,
Controlling error orientation to improve quantum algorithm success rates.
{\it Phys. Rev. A} {\bf 99}, 032318 (2019).

\bibitem{Broadband}
J. T. Merrill, K. R. Brown,
Progress in compensating pulse sequences for quantum computation.
\href{https://arxiv.org/abs/1203.6392}{https://arxiv.org/abs/1203.6392} (2012).

\end{thebibliography}
\end{document}